\documentclass[a4paper,fleqn]{cas-dc}

\usepackage[numbers]{natbib}
\usepackage{algorithm}
\usepackage{algpseudocode}
\usepackage{bbm}
\usepackage{diagbox}

\def\tsc#1{\csdef{#1}{\textsc{\lowercase{#1}}\xspace}}
\tsc{WGM}
\tsc{QE}
\tsc{EP}
\tsc{PMS}
\tsc{BEC}
\tsc{DE}

\begin{document}
\let\WriteBookmarks\relax
\def\floatpagepagefraction{1}
\def\textpagefraction{.001}

\shorttitle{Generalizable Agent Modeling for Agent 
Collaboration-Competition Adaptation}

\shortauthors{Chenxu Wang et~al.}

\title [mode = title]{Generalizable Agent Modeling for Agent 
Collaboration-Competition Adaptation with Multi-Retrieval and Dynamic Generation}                      





%







\affiliation[1]{organization={Beijing University of Posts and Telecommunications},
    city={Beijing},
    postcode={100876}, 
    country={China}}

\author[1]{Chenxu Wang}[style=chinese]
\fnmark[1] 
\ead{wangchenxu@bupt.edu.cn} 

\author[1]{Yonggang Jin}[style=chinese]
\fnmark[1] 
\ead{daze@bupt.edu.cn} 

\author[1]{Cheng Hu}[style=chinese]
\fnmark[1] 
\ead{hucheng@bupt.edu.cn} 



\affiliation[2]{organization={Polixir},
    city={Nanjing},
    country={China}}

\author[2]{Youpeng Zhao}[style=chinese]
\fnmark[2] 
\ead{youpeng.zhao@polixir.ai} 

\affiliation[3]{organization={Beijing Institute of Technology},
    city={Beijing},
    postcode={100081}, 
    country={China}}
\author[3]{Zipeng Dai}[style=chinese]
\fnmark[3] 
\ead{daizipeng@bit.edu.cn} 

\author[2]{Jian Zhao}[style=chinese]
\fnmark[2] 
\ead{jian.zhao@polixir.ai} 

\affiliation[5]{organization={Tsinghua University},
    city={Beijing},
    postcode={100084}, 
    country={China}}
    
\author[5]{Shiyu Huang}[style=chinese]
\fnmark[5] 
\ead{huangsy1314@163.com} 

\author[1]{Liuyu Xiang}[style=chinese]
\fnmark[1] 
\ead{xiangly@bupt.edu.cn} 

\affiliation[4]{organization={Institute of Automation, Chinese Academy of Sciences},
    city={Beijing},
    postcode={100190}, 
    country={China}}
\author[4]{Junge Zhang}[style=chinese]
\fnmark[4] 
\ead{jgzhang@nlpr.ia.ac.cn} 

\author[1]{Zhaofeng He}[style=chinese]
\fnmark[1] 
\ead{zhaofenghe@bupt.edu.cn} 
\cormark[1]

\cortext[cor1]{Zhaofeng He}



\begin{abstract}
Adapting a single agent to a new multi-agent system brings challenges, necessitating adjustments across various tasks, environments, and interactions with unknown teammates and opponents. 
Addressing this challenge is highly complex, and researchers have proposed two simplified scenarios, Multi-agent reinforcement learning for zero-shot learning and Ad-Hoc Teamwork. 
Building on these foundations, we propose a more comprehensive setting, Agent Collaborative-Competitive Adaptation (ACCA), which evaluates an agent to generalize across diverse scenarios, tasks, and interactions with both unfamiliar opponents and teammates. 
In ACCA, agents adjust to task and environmental changes, collaborate with unseen teammates, and compete against unknown opponents. 
We introduce a new modeling approach, Multi-Retrieval and Dynamic Generation (MRDG), that effectively models both teammates and opponents using their behavioral trajectories. 
This method incorporates a positional encoder for varying team sizes and a hypernetwork module to boost agents' learning and adaptive capabilities. 
Additionally, a viewpoint alignment module harmonizes the observational perspectives of retrieved teammates and opponents with the learning agent. 
Extensive tests in benchmark scenarios like SMAC, Overcooked-AI, and Melting Pot show that MRDG significantly improves robust collaboration and competition with unseen teammates and opponents, surpassing established baselines. 
Our code is available at: https://github.com/vcis-wangchenxu/MRDG.git
\end{abstract}


\begin{highlights}
\item This research mainly focuses on the generalization performance of a single agent in long-sequence decision-making tasks. 
Its goal is to endow the agent with the ability to dynamically adapt to different tasks and to interact with agents adopting different strategies. 
The challenges faced by this research lie not only in endowing the agent with the ability to dynamically adapt to tasks, but more importantly, in enabling it to possess the ability to dynamically interact with agents with diverse strategies;

\item We contend that the extant research on the generalization of intelligent agents is incomplete. Hence, we put forward a more comprehensive framework for investigating the generalization of intelligent agents. In this framework, by means of dynamic retrieval, we dynamically generate the parameters of the intelligent agent's policy network. This endows the agent with the capacity to dynamically adapt to tasks and the strategies of other intelligent agents;

\item To comprehensively validate the effectiveness of the proposed method, we executed experiments across multiple long-sequence decision - making simulation environments, namely SMAC, Overcooked-AI, and Melting Pot. Additionally, a comparative analysis was performed against other extant research achievements in the domain of agent generalization. The results vividly illustrate that the method we put forward holds substantial performance advantages;

\item We propose a novel agent modeling method called Multi-Retriever and Dynamic Generation (MRDG), which constructs the learner’s knowledge base by integrating the actions and attributes of agents within a multi-agent system (MAS). 
This approach enhances the learner’s adaptability to dynamic scenarios and facilitates improved generalization in evaluation scenarios;

\item On the theoretical front, this research contributes to a profounder comprehension of the learning mechanisms and decision-making processes of agents, thereby enriching and refining the fundamental theoretical framework of artificial intelligence. From an application standpoint, a single agent equipped with the capabilities to adapt dynamically to diverse tasks and engage in interactions with other agents holds great promise for extensive applications across a wide spectrum of fields.

\end{highlights}


\maketitle

\section{Introduction}

Adapting independent agents to new Multi-Agent Systems (MAS)~\cite{van2008multi, dorri2018multi, balaji2010introduction} poses formidable challenges. Agents must adapt not only to diverse tasks and complex environments but also to interactions with unknown teammates and opponents. 
This necessitates significant flexibility in dynamic decision-making and sophisticated strategic learning capabilities. 
In contrast to single-agent systems, which typically assume environmental stationarity (i.e., fixed state transition probabilities and reward functions), multi-agent scenarios are characterized by fundamental non-stationarity. 
The dynamically updating strategies of other agents continuously alter the interaction dynamics, leading to an exponential expansion of the unknown hypothesis space faced by an individual agent. 
Specifically, treating other agents' strategies as latent environmental variables means the joint strategy space requiring modeling grows combinatorially with the number of agents. 
This combinatorial explosion often renders the adaptation problem intractable for traditional reinforcement learning methods. 
Their reliance on the stationary Markov Decision Process (MDP) framework is ill-suited for capturing the environmental non-stationarity induced by dynamic strategic interactions.

In the StarCraft Multi-Agent Challenge (SMAC), for instance, varying combinations of enemy unit strategies, such as target selection and formation changes, can generate numerous distinct interaction patterns. 
Treating these diverse strategies merely as environmental noise hinders an agent's ability to generalize effectively from limited training data when using traditional methods. 
Successfully addressing this adaptation challenge is crucial for fields including automated control~\cite{ge2023communication, afrin2021resource} and game AI~\cite{lin2023tizero, xu2023noncooperative}. 
For example, in robot swarm collaboration, newly added agents must quickly adapt to the heterogeneous strategies of existing members; similarly, in real-time strategy games, AI must dynamically counter the unknown tactics employed by human players or opposing AIs.

\begin{figure*}[!ht]
\centering     
\includegraphics[width=0.9\textwidth]{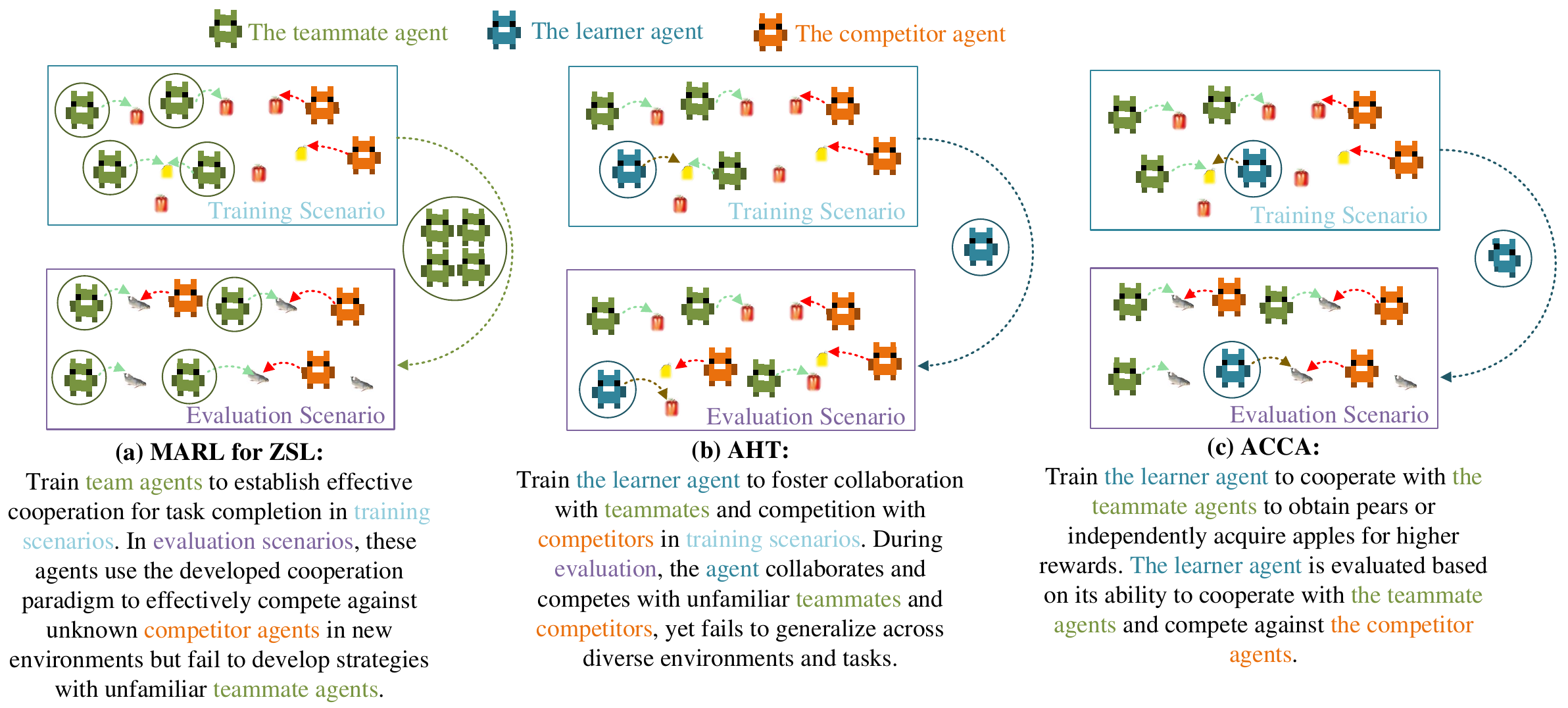}
\caption{Comparison between existing frameworks for validating the generalization performance of intelligent agents and our proposed ACCA. 
\textbf{Left:} In MARL for ZSL, the focus is on training teams of intelligent agents for versatility across varied tasks and scenarios, while the cooperative aspect of the teams with unfamiliar teammates is overlooked. 
\textbf{Middle:} In AHT, the emphasis is on training an agent to effectively cooperate with unknown teammates on specific tasks, yet it fails to address generalization across different tasks and scenarios. 
\textbf{Right:} In ACCA, there is a strong emphasis not only on generalizing to diverse scenarios and tasks but also on adapting to the evolving strategies of both teammates and opponents.}
\label{fig1}
\end{figure*}

To evaluate the generalization performance of adaptable agents in complex MAS scenarios, distinct training and testing environments were established, as depicted in Figure~\ref{fig1}. \textbf{Training Scenario:} In a representative fruit collection task, adaptable agents can acquire rewards through two primary methods: individual apple collection, yielding a smaller reward, or cooperative pear collection with teammates, resulting in a larger reward. Given the fixed quantity of fruit within this scenario, agents must also engage in competitive interactions with opponents. \textbf{Testing Scenario:} To simulate realistic unknown conditions, the testing scenario incorporates novel variations in the environment, task objectives, teammate strategies, and opponent strategies. These specific configurations were not encountered by the agents during the training phase.

\begin{table*}[htbp]
 \centering
 \caption{Symbols Description}
 \begin{tabular}{cc}\toprule
    \multicolumn{1}{c}{Symbol} & \multicolumn{1}{c}{Connotation}
    \\\cmidrule(lr){1-2}
    $\Psi$   & Diversity Policy Pool (DPP) \\
    $C$   & The set of coordination scheme \\
    $G^1$ & A controllable agent (the learner) \\
    $G^{-1}$ & A group of uncontrollable agents (teammates or competitors) \\
    $\mathbf{a}^{-1}$ & The joint action of uncontrollable agents \\
    $SE$ & The substrate environment \\
    $\langle \pi_{\theta_i}^{-1} \rangle_{i=2}^N$ & The joint policies of uncontrollable agents \\
    $\pi_{\theta}^1$ & The policy of the learner \\
    $D$  &  Replay Buffer \\
    $D_r$ & Episodic Memory \\
    $VA$ &  Viewpoint Alignment Encoder \\
    $o_t^i$ & The observations of the $i$-th teammates and competitors at time $t$ \\
    $x_t^i$ & The features encoded by the VA Encoder for the $i$-th teammates and competitors at time $t$ \\
    $\theta^{marl}$ & The parameters of the policy network obtained by the learner through MARL algorithm training \\
    $\theta^{hype}$ & The parameters of the policy network obtained by the learner through the hypernetwork \\
    $\theta_i^V$ & The network parameters of the VA Encoder corresponding to the i-th teammates and competitors \\
    $\theta^{retr}$ & The parameters of retrieval network \\
    $\theta^{hy}$ & The parameters of hypernetwork \\
    $\theta^{RI}$ & The parameters of Re-initialization\\
    \\\bottomrule
 \end{tabular}
 \label{sign}
\end{table*}

Currently, the primary approaches to address the challenge of agent adaptability in novel multi-agent scenarios include Multi-Agent Reinforcement Learning for Zero-Shot Learning (MARL for ZSL) and Ad Hoc Teamwork (AHT). MARL for ZSL~\cite{ding2024multi, guo2024heterogeneous, xue2022heterogeneous} aims to train a group of agents to collaborate effectively and complete tasks in unfamiliar environments for which they have not been pre-trained. As illustrated in Figure~\ref{fig1} (left), during the training phase, team agents might collect pears through internal cooperation to obtain higher rewards, while also having the option to collect apples individually for lower rewards and compete with opponents. Subsequently, these agents are deployed in a new environment where they compete with opponents to collect fish for rewards. However, this approach primarily focuses on the generalization of team agents across different tasks and scenarios, often overlooking the development of their collaborative capabilities when interacting with unfamiliar teammates. In practical applications, when confronted with newly added teammates whose strategies are unknown, these agents typically struggle to rapidly adjust their collaborative methods, resulting in diminished cooperation efficiency.

AHT~\cite{stone2010ad}, conversely, endeavors to enable individual agents to collaborate effectively with diverse, unfamiliar teammates across various tasks in a zero-shot manner. As depicted in Figure~\ref{fig1} (middle), during training, the learner agent cooperates with teammates to collect pears for high rewards while simultaneously competing with opponents. Subsequently, this learner agent is placed in an unfamiliar environment where it must cooperate with unknown teammates and compete against unfamiliar opponents. However, AHT predominantly concentrates on the collaboration between an agent and unknown teammates within a specific task, often neglecting the generalization problem across different tasks and environmental contexts. In real-world scenarios, tasks and environments are frequently complex and varied. Agents trained using AHT methods often struggle to maintain optimal performance when faced with new tasks and environments, primarily due to insufficient adaptability to environmental changes and a limited capacity for cross-task strategy adjustment.

To enhance the adaptability and generalization~\cite{reed2022generalist} capabilities of agents, we introduce the Agent Collaboration-Competition Adaptation (ACCA) framework. As shown in Figure~\ref{fig1} (right), ACCA is designed to enable agents to respond flexibly to dynamic shifts in collaboration and competition within novel scenarios, even when the strategies of allies and opponents are initially unclear, thereby facilitating efficient adaptation. The ACCA framework requires agents to collaborate with unknown teammates and compete against unknown opponents across a range of different tasks. We posit that ACCA presents a new research direction for agent generalization. By developing agents capable of adapting to unfamiliar environments, tasks, and strategies, we advance progress towards creating truly intelligent, adaptable, and broadly generalizable agents.

This paper aims to train an agent that generalizes well, excelling in both cooperation and competition across multi-agent systems, while adapting to variations in strategies and the number of teammates and adversaries. In the ACCA framework, we introduce a generalizable agent modeling method, Multi-Retriever and Dynamic Generation (MRDG) assists learners in adapting to the strategic and numerical changes of other intelligent agents.
We create a Diversity Policy Pool (DPP) to store various strategies from teammates and adversaries, enriching the learner's policy network. Episodic memories are built for both sides, allowing the model to explicitly represent this data. Through efficient retrieval, the learner captures high-reward behaviors from teammates and low-reward behaviors from adversaries. Positional encoding helps distinguish agent attributes.
A hypernetwork module adjusts the policy network parameters based on behavioral data to ensure robust cooperation with teammates. Additionally, a Viewpoint Alignment (VA) module aligns teammates’ and adversaries’ observations with the learner’s, improving retrieval effectiveness in new scenarios.

To assess the generalization of the trained agent (the learner), we select diverse benchmarks as validation platforms and conduct extensive experiments on SMAC~\cite{samvelyan19smac} and Overcooked-AI~\cite{carroll2020utility} (pure common interest), and Melting Pot~\cite{leibo2021scalable} (mixed motive). 
These benchmarks encompass a variety of substrates across diverse scenarios and tasks, thereby providing a robust means of validating the efficacy of our method within the proposed setting.
The experimental results demonstrate that MRDG significantly enhances the generalization of cooperative behaviors when collaborating with teammates and competing against adversaries, both under conditions of undefined strategies and quantities. 

Our main contributions are as follows:
\begin{enumerate}
\itemsep=0pt
\item We introduce a comprehensive framework, the Agent Collaboration-Competition Adaptation (ACCA), which requires agents to demonstrate generalizability across diverse scenarios and tasks, as well as the ability to collaborate and compete with unfamiliar teammates and opponents. 
This framework exhibits great potential in advancing research on general agents;
\item We propose a novel agent modeling method called Multi-Retriever and Dynamic Generation (MRDG), which constructs the learner’s knowledge base by integrating the actions and attributes of agents within a multi-agent system (MAS). 
This approach enhances the learner’s adaptability to dynamic scenarios and facilitates improved generalization in evaluation scenarios;
\item Experimental results indicate that MRDG outperforms the current state-of-the-art baselines in the proposed setting. 
This finding underscores the significant advantages of our method in addressing generalization challenges involving unknown scenarios, tasks, and the strategic behaviors of both teammates and adversaries;
\end{enumerate} 

\section{Related work}

\subsection{Multi-Agent Reinforcement Learning for Zero-Shot Learning}
MARL for ZSL~\cite{pourpanah2022review, mancini2021open, keshari2020generalized} investigates the dynamics within populations of intelligent agents that can perform well without re-training on new tasks or environments. 
The collaboration among these agents is essential for improving their adaptability and generalization skills. 
Existing methods include meta-learning~\cite{hu2021distributed, yun2023quantum, munir2021multi, verma2020meta, liu2021task}, transfer learning~\cite{shi2021lateral, yang2021efficient, wu2022strategic} and generative adversarial networks (GANs)~\cite{kumar2020harnessing} for MARL~\cite{zhang2021multi, yang2020overview, padakandla2021survey}, etc.
Ding and Zheng introduce a cross-domain zero-shot learning 
(CDZSL) approach~\cite{ding2024multi} for MAS, which significantly reduces the adaption time for agents to complete new cooperative tasks without prior examples. 
Similarly, Xue et al. present a novel coevolutionary strategy~\cite{xue2022heterogeneous} to tackle the heterogeneous zero-shot coordination challenge in cooperative MARL problems, with experimental validation across diverse tasks. 
Zhao et al. propose a neural network-based pre-trained model~\cite{zhao2023towards} that achieves general zero-shot capability on a diverse array of previously unseen restless multi-arm bandit problems, addressing limitations of prior research and accommodating the dynamic nature of real-world applications.
However, these methods primarily focus on intra-agent cooperation and do not address the capability to collaborate with newly introduced agents.

\subsection{Ad Hoc Teamwork} 
Ad hoc teamwork (AHT) aims to create autonomous agents capable of collaborating with previously unknown teammates on various tasks~\cite{stone2010ad, mirsky2022survey}. 
Classic methods include type-based~\cite{lupu2021trajectory} approaches and experience replay~\cite{chen2020aateam} to characterize domain knowledge and the use of planning~\cite{sarratt2015tuning} and meta-learning~\cite{zintgraf2021deep} for action selection. 
Recent work introduces the CSP~\cite{ding2023coordination} approach, which employs a disentangled scheme probing module to represent and classify new teammates beforehand using limited pre-collected episodic data, thereby facilitating multi-agent control. 
This method emphasizes studying an agent's collaboration with unknown teammates to complete tasks, focusing on the diversity of teammates' strategies and cooperation methods. 
However, it overlooks changes in the environment and tasks, and lacks generalization ability to new environments and tasks.
We adhere strictly to the ACCA framework and use retrieval methods to maintain the learner’s adaptability to new scenarios.

\subsection{Agent Modeling} Agent modeling~\cite{albrecht2018autonomous} approaches aim to provide auxiliary information by modeling the behaviors of teammates, which includes insights such as teammates' goals or future actions, to aid in decision-making. 
MeLIBA~\cite{zintgraf2021deep} conditions the ad hoc agent’s policy on a belief about its teammates, updated according to the Bayesian rule. 
Additionally, several studies explore the generation of diverse agent policies~\cite{eysenbach2018diversity, chen2022dgpo, huang2022vmapd, park2023metra}, enhancing the training of ad hoc agents~\cite{canaan2019diverse}. 
A recent work, ODITS~\cite{gu2021online}, introduces an information-based regularizer to automatically approximate the hidden state of the global encoder with that of the local encoder. 
RPM~\cite{qiu2022rpm} facilitates MARL by storing and ranking policies based on training episode returns, enabling agents to select diverse behavioral strategies during training, thereby enhancing the generalizability of policies and the ability to interact with previously unseen agents. 
However, these methods implicitly model teammates and heavily rely on training data to infer their behaviors and strategies. 
Consequently, the learner may focus on cooperating with familiar teammates and overlook new ones when they appear.

RPM and ODITS largely overlook opponent modeling, limiting their adaptability to shifting strategies in competitive scenarios where inability to predict opponent actions severely hampers counter-strategy. MRDG, conversely, models both teammates and opponents, enabling more effective adaptation to dynamic environments and proficient strategy adjustment against unknown opponents by analyzing their behavioral patterns, thus enhancing competitive survival and success.

RPM, while storing diverse strategies for generalization, shows limited adjustment capability against frequent teammate strategy changes due to its static, pre-stored library. CSP, though identifying new teammates via a disentangled module, lacks flexibility with task/environmental variations, prioritizing teammate identification over dynamic environmental considerations. MRDG excels in adaptability through dynamically generated strategy network parameters, instantly creating suitable strategies based on real-time scenarios and agent behavior, independent of a fixed library, thus better coping with new scenarios and strategic shifts.

Furthermore, methods like ODITS inadequately address inherent inter-agent viewpoint differences arising from factors like position, leading to disparate information access. MRDG's viewpoint alignment module maps teammate and opponent viewpoints to the learner’s perspective, significantly enhancing retrieval effectiveness and information comprehension in novel scenarios.

\section{Preliminaries}
\subsection{Multi-Agent Reinforcement Learning} We define the problem as a decentralized partially observable Markov decision process (Dec-POMDP)~\cite{oliehoek2016concise}, delineated by $\mathcal{G}=\langle \mathcal{N}, \mathcal{S}, \{ \mathcal{O}^i \}_{i=1}^{N},  \{ \mathcal{A}^i \}_{i=1}^{N}, \Omega, P, R, \gamma, \rho \rangle$. 
Within this framework, $\mathcal{N}=\{1, \dots, N\}$ represents the agents, $\mathcal{S}$ means the global states, $\mathcal{O}^i$ is the observation set of agent $i$, and $\mathcal{A}^i$ denotes the corresponding action set. 
The discount factor, $\gamma$, ranges from 0 to 1, and $\rho: \mathcal{S} \rightarrow [0, 1]$ specifies the initial state distribution. 
At each time step, agent $i \in \mathcal{N}$ acquires a local observation $o^i \in \mathcal{O}^i$ via the observation function $\Omega(s, i)$ and chooses an action $a^i \in \mathcal{A}^i$ based on its individual policy $\pi_{\theta_i}(a^i|\tau^i)$, where $\tau^i$ captures the agent’s historical sequence of observations and actions. 
The agents' collective action $\mathbf{a}=\langle a^1_t, \dots, a^n_t \rangle$ transitions the system to the subsequent global state $s'$ following the transition function $P(s' | s, \mathbf{a})$. 
The reward function is constructed as $R=\times_{i=1}^{N} r^i$, where $r^i$ denotes the reward for agent $i$ contingent upon the state and joint actions. 
Each agent endeavors to maximize its return through reinforcement learning (RL), refining its policy $\pi_{\theta_i}$ to optimize the objective outlined by:
\begin{equation}
    \mathcal{J}(\pi_{\theta_i}) \triangleq \mathbb{E}_{s_{0:\infty} \sim \rho^{\mathcal{G}}_{0:\infty}, a_{0:\infty}^{i} \sim \pi_{\theta_i}} \left[ \sum_{t=0}^{\infty} \gamma^{t}r_{t}^{i} \right],
\label{Jpii}
\end{equation}
where $\mathcal{J}(\pi_{\theta_i})$ evaluates the policy performance of RL methods~\cite{fujimoto2018addressing}.

\subsection{Agent Distinction in MAS} Our objective is to develop an intelligent agent capable of establishing effective cooperation with unfamiliar teammates and engaging in competitive strategies against unknown adversaries across a variety of tasks and scenarios. 
Without loss of generality, we categorize agents within a MAS into two groups to achieve this: a controllable agent $G^1 \subseteq \mathcal{N}$ and uncontrollable agents $G^{-1} \subseteq \mathcal{N}$. 
Here, $G^1$ represents the agent undergoing training, whereas $G^{-1}$ includes the uncontrollable agents that $G^1$ must adapt to. 

We denote the observation, action and policy for $G^{1}$ as $o^1, a^1, \pi^1$, respectively, and the joint observation, action and policy for $G^{-1}$ are represented using $\mathbf{o}^{-1}=\langle o^i \rangle_{i=2}^N$, $\mathbf{a}^{-1}=\langle a^i \rangle_{i=2}^N$, $\mathbf{\pi}^{-1}=\langle \pi_{\theta_i} \rangle_{i=2}^N$. 
Consequently, the joint team policy is denoted as $\mathbf{\pi}=\langle \mathbf{\pi}^1, \mathbf{\pi}^{-1} \rangle$.  
Let $\prod_f$ represent the adaptability of the agent and define the substrate environment as $SE$, where different $SE$s correspond to different tasks. 
For simplicity, we associate each $SE$ with a specific task.
We define $C$ as the set of multi-agent interaction policies, with each $c_i$ representing a distinct set of joint policies such that $c_i \cap c_j=\emptyset$ for $i \neq j$. 
The goal of MARL for ZSL is to train a joint policy $\pi$ that can still perform well on new $SE$s without further training, disregarding the changes in $C$. 
The aim of AHT is to train an agent with a policy $\pi^1$ that can effectively cooperate with unknown $c$, which is part of $C$, on the same $SE$, disregarding the changes in $SE$. 
To be noted, AHT only considers interactions in purely cooperative scenarios and does not account for mixed-motive scenarios.
Our objective is to comprehensively consider the changes in $C$ and $SE$ to achieve true generalization for a single agent, such that $\pi^1 \in \prod_f$.

\subsection{Problem Formalization} We intend to develop an autonomous agent $G^1$, referred to as the learner agent, that coordinates and competes with a non-controllable group $G^{-1}$ under any given joint policies $c \in C$ in unknown environment $SE$. 
We presuppose that $G^{-1}$ is incapable of adaptation, adhering strictly to a fixed scheme regardless of the actions of $G^1$. 
Our ultimate goal is to maximize team rewards. 
To this end, we postulate a shared reward structure where the learner agent's optimal policy $\pi_{\theta}^{1*}$ is designed to maximize the joint action value $Q^{\pi^{1}}(s, a^1, \mathbf{a}^{-1})$. 
This value represents the expected cumulative team reward across various cooperation and competition scenarios:
\begin{equation}
    \begin{aligned}
    Q^{\pi^{1}}(s,a^1, \mathbf{a}^{-1})= & \mathbb{E}_{a_{t=0:\infty}^{1} \sim \pi^{1},\mathbf{a}_{t=0:\infty}^{-1} \sim \mathbf{\pi}^{-1},  \mathbf{\pi}^{-1} \in c} \\ & \left[ \sum_{t=0}^{\infty} \gamma^{t} r_t |P, \rho \right],
    \end{aligned}
\label{CompQ}
\end{equation}
\begin{equation}
    Q^{\pi^{1*}}(s,a^1, \mathbf{a}^{-1}) \ge Q^{\pi^{1}}(s,a^1, \mathbf{a}^{-1}), \forall {\pi^{1}, s, a^1,\mathbf{a}^{-1}}.
\label{QStart}
\end{equation}
The optimal policy $\mathbf{\pi}_{\theta}^{1*}$, parameterized by $\theta$, for $G^1$ seeks to maximize the discounted cumulative reward, defined as $\theta^{*} = \mathop{\arg\max}\limits_{\theta} Q^{\pi_{\theta}^{1}}(s,a^1, \mathbf{a}^{-1})$.

\begin{figure*}[!ht]
\centering
\includegraphics[width=0.8\textwidth]{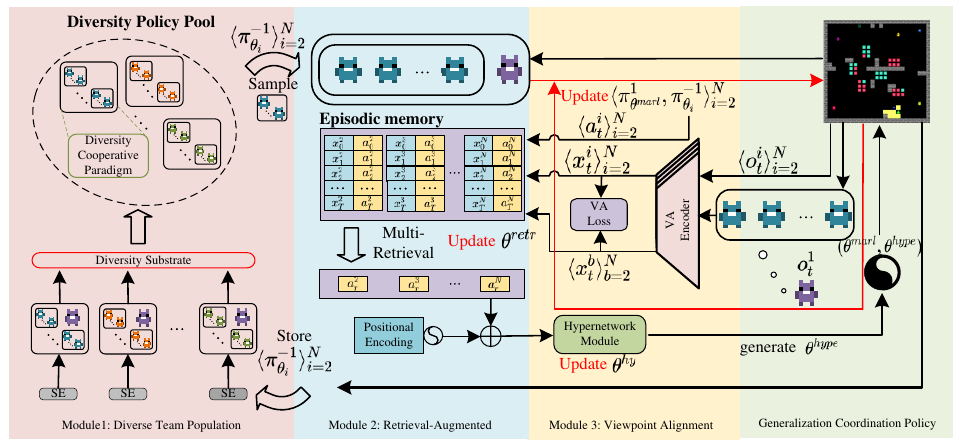} 
\caption{\textbf{Schematics of MRDG and core mechanism for adaptive generalization in MAS}. This illustrates how a \textbf{Diverse Team Population module} establishes policy diversity, a \textbf{Retrieval-Augmented Agent Modeling module} extracts agent features and structure, a \textbf{Viewpoint Alignment module} unifies observations, and a \textbf{HyperNetwork module} dynamically generates policy parameters. This creates the "sample-interact-retrieve-generate-store" closed loop, enabling rapid adaptation to varying agent policies, numbers, and observation viewpoints to address generalization challenges.}
\label{fig2}
\end{figure*}

\section{Methodology}
This section describes how the MRDG framework addresses issues of the learner achieving effective cooperation with teammates and competition with opponents in an end-to-end manner, as illustrated in Figure~\ref{fig2}. 
MRDG adapts to unfamiliar substrates, tasks, and policies of teammates and opponents using three primary modules: 
(1) \textbf{Diverse Team Population:} Various substrates, tasks and agent strategies are sampled from the training environment to enhance the diversity of teammate strategies. 
(2) \textbf{Retrieval-Augmented Teammate and Adversary Modeling:} Teammate and adversary data representations are enriched through retrieval, which informs part of the learner’s network parameters, synthesized by the Hypernet. 
(3) \textbf{Viewpoint Alignment:} This module maps the third-person perspective of teammates and adversaries to the first-person perspective of the learner, facilitating adaptation to new situations.

We partition the agents in the multi-agent system into group $G^1$, the agent designated for training, and group $G^{-1}$, encompassing other agents (potential teammates or opponents) to which $G^1$ must adapt. 
The controllability of $G^{-1}$ differs significantly between the training and testing phases.

\textbf{During training}, the strategies employed by $G^{-1}$ are \textbf{controlled} by the training process. 
To foster generalization capabilities in $G^1$, we train the agents using MARL algorithms. 
Specifically, DPP is used to dynamically generate diverse strategies for $G^{-1}$ (including both cooperative and competitive approaches). 
Strategy combinations from the DPP are randomly sampled in each training episode, compelling $G^1$ to adapt to varied interaction patterns. 
The primary objective of controlling $G^{-1}$'s strategies during training is to create diverse interaction scenarios, thereby preventing $G^1$ from overfitting to a limited set of behaviors. 
Conversely, \textbf{during the testing phase}, $G^{-1}$'s strategies are considered \textbf{non-controllable}: they are fixed throughout the test and represent policies potentially unseen by $G^1$ during training. 
These test strategies are drawn from a predefined set, potentially including policies from the DPP held out from the training set. 
This setup simulates real-world conditions where $G^1$ must dynamically adapt its strategy by observing $G^{-1}$'s behavior, leveraging components such as a Retrieval Network and a Hypernetwork.

\begin{algorithm}[htb]
\caption{The MRDG Training Process}
\label{alg:algorithm2}
\textbf{Input}: Initialize $\pi_{\theta}^1$, $\langle \pi_{\theta_i}^{-1} \rangle_{i=2}^N$, $\Psi$, $\theta^{hy}$, $\theta^{retr}$, $\langle \theta_i^V \rangle_{i=2}^N$, $\lambda$, $\gamma$, $D$, $D_r$, $SE$, $SE^{'}$, $c$, $c^{'}$ and $\theta^{RI}$;\\
\textbf{Input}: Initialize the policy of learner: $\pi_{\theta}^1$ := $\pi_{[\theta^{marl}; \theta^{hype}]}^1$;
\begin{algorithmic}[1] 
\For{$e$ in $1, \dots,$ MAX\_EPISODE}
\If{$\Psi$ sampling}
\State $SE$ $\gets$ Sampling($\Psi$);
\State $\langle \pi_{\theta_{i}}^{-1} \rangle$ $\gets$ $c_i:$ Sampling($SE$);
\EndIf
\For{$t$ in $1, \dots,$ MAX\_TIMESTEP}
    \State $D$ $\gets$ GatherTrajectories($\langle \pi_{\theta}^1, \pi_{\theta_i}^{-1}\rangle, c$);
    \For{$i$ in $2, \dots,$ MAX\_TEAMMATES}
        \State $D_r$ $\gets$ GatherData $(x_{t}^{i}, a_{t}^{i})$ using $o_t^i$ and $f_{\theta_i^V}$;
        \State Compute $\mathcal{C}_{i}$ using $x_{t}^{i}$ and $x_t^b$;
        \State $\theta_{i}^{V}$ $\gets$ update($\mathcal{C}_{i}$, $\theta_{i}^{V}$);
        \State Choose $m$ actions for agent $i$ using $f_{\theta^{retr}}$ 
        \State $\qquad$ and $x_t^b$;
        \State Get RetrievalAction $a_r^{i}$; 
    \EndFor
    \State Compute $\vec{p_i}^{k}$;
    \State Generate $\pi_{\theta^{hype}}^1$ using $\vec{p_i}^{k}$, $a_r^i$ and $\theta^{hy}$;
\EndFor
\State Compute $\mathcal{L}(\theta_i)$ using $D$;
\State $\langle \pi_{\theta^{marl}}^1$, $\pi_{\theta}^{-1} \rangle$ $\gets$ MARLTraining($\langle \pi_{\theta^{marl}}^1$,$\pi_{\theta}^{-1} \rangle$,$D$);
\State $\theta^{hy}$ $\gets$ Update($\theta^{hy}$, $\mathcal{L}(\theta_i)$);
\State $\theta^{retr}$ $\gets$ Update($\theta^{retr}$, $\mathcal{L}(\theta_i)$);
\If{$\theta^{marl}$ Re-initialization}
    \State $\theta^{marl} = \lambda \theta^{marl} + \gamma \theta^{RI}$;
\EndIf
\State $\Psi$ $\gets$ Update($\langle \pi_{\theta_{i}}^{-1} \rangle, \Psi, SE_c$);
\State $\overline{R}$ $\gets$ Evaluate($\pi_{\theta}^1, SE^{'}_{c^{'}}$);
\EndFor
\end{algorithmic}
\textbf{Output}: $\pi_{\theta}^1$
\end{algorithm}

\subsection{Overview}
Our objective is to train an agent capable of generalizing across diverse environments, tasks, and team compositions, and adept at forming both collaborative and competitive interactions. 
The agent's policy network parameters are denoted by $\theta = \langle \theta^{marl}, \theta^{hype} \rangle$.

Initially, we randomly select a substrate and its corresponding optimization objective from the Diverse Team Population. 
Subsequently, we sample the strategies of team agents from this substrate and assign them accordingly.
Under this framework, the learner and team members’ strategies are co-optimized using the MARL algorithm to refine the learner’s policy network parameters, $\theta^{marl}$.

Throughout the optimization process, the observations and actions of team members, transformed via Viewpoint Alignment, are recorded in the Episodic Memory.
This alignment adjusts team members' perspectives to that of the learner.
Using features processed from the learner's observations via Viewpoint Alignment, we retrieve the Episodic Memory for the closest match to the team member’s observation-action pairs, which facilitates explicit modeling of team behaviors. 
These retrieved action values are then processed through Positional Encoding and fed into the Hypernetwork Model, which outputs the learner’s policy parameters, $\theta^{hype}$. 

Ultimately, the refined policy parameters of the team members are cataloged in the Diverse Team Population under their respective substrates.
The Pseudocode for this approach is outlined in Algorithm~\ref{alg:algorithm2}, with detailed descriptions of the modules provided below.

\subsection{Diverse Team Population} 
To ensure effective collaboration with diverse teammates, it is critical to expose the learner to a wide range of teammate strategies during training. 
Inspired by RPM, we establish a Diversity Policy Pool (DPP), denoted as $\Psi$, comprising various substrates and policies. 
We define the substrate environment as $SE$, with different joint policies $c_i$. 
As the learner interacts with different teammates across these substrates, the training episode yields a return $R$ for all agents, governed by policies $\langle \pi_{\theta}^1, \pi_{\theta_i}^{-1} \rangle_{i=2}^{N}$, where $\pi_{\theta}^1$ represents the learner’s strategy, and $\langle \pi_{\theta_i}^{-1} \rangle_{i=2}^{N}$ denotes the teammates’ joint strategies. 
For simplicity, $\langle \pi_{\theta_i}^{-1} \rangle_{i=2}^{N}$ is abbreviated to $\langle \pi_{\theta}^{-1} \rangle$. 
During the interaction, each policy’s $Q$ value $Q_i$ is optimized by minimizing the following regression loss with TD-learning:
\begin{equation}
\begin{split}
\mathcal{L}(\theta_{i}) \triangleq \mathbb{E}_{(o_{t}^{i},a_{t}, \mathbf{a}_{t}^{-1}, r_t) \sim D} 
\left[  \left(  y_{t}^{i}-Q_{\theta_i}^{i} (o_t^{i}, a^{1}_{t}, \mathbf{a}^{-1}_{t})  \right)^2   \right],
\end{split}
\label{CompLossmarl}
\end{equation}
where $y_{t}^{i}=r_{t}^{i}+\gamma \max_{\mathbf{a}^{'}} Q_{\overline{\theta_i}}^{i}(o_{t+1}^{i}, a^{i, '}, \mathbf{a}^{-1'}) $ and $\theta_i$ represents the parameters of the agents. $\overline{\theta_i}$ is the parameters of the target $Q^i$ and periodically copied from $\theta_i$. $(o_{t}^{i},a_{t}, \mathbf{a}_{t}^{-1}, r_t)$ is a sample from the replay buffer $D$. 

To maintain and enhance the diversity within the DPP, we implement dynamic interactive storage. 
Before each episode begins, a substrate $SE$ is sampled from $\Psi$, followed by a random selection of a joint policy $c$ in multi-agent interaction policies $C$ from the corresponding memory slot of $SE$. 
During training, $\langle \pi_{\theta^{marl}}^1, \pi_{\theta}^{-1} \rangle$ is updated according to the objectives in $c$, where $\pi_{\theta^{marl}}^1 \in \pi_{\theta}^1$. 
Focusing on diversifying teammates’ policies, at the end of the episode, the joint strategies used during training are saved in the corresponding $c$ of the substrate in $\Psi$, as $c_i: \langle \pi_{\theta}^{-1} \rangle$ are stored in $\Psi$ by adding them to the corresponding $c$ of $SE$ memory slot, $\Psi[SE_{c}].add({ \pi_{\theta}^{-1} })$. 
Subsequently, strategies for all teammate agents are refreshed with newly sampled strategies from $\Psi$, ensuring a diversity of multi-agent trajectories.

The adoption of TD-learning is principally motivated by its synergy and compatibility with the dynamic interaction characteristics inherent in MAS and the random sampling mechanism of the DPP. 
As the DPP module randomly samples substrates and agent strategies during training, the interaction trajectories experienced by agents exhibit strong non-stationarity. Consequently, Generalized Advantage Estimation (GAE), which depends on the temporal correlation of continuous trajectories, encounters difficulties in accurately estimating the advantage function under conditions of rapid policy switching and is susceptible to high variance.

In contrast, TD-learning, with its single-step bootstrapping mechanism, updates the value function using only the state-action pairs from the current and subsequent timesteps, thus demanding minimal trajectory integrity. 
This facilitates real-time online learning in non-stationary environments. 
This property naturally synergizes with the Retrieval-Augmented module within the MRDG framework. When the Retrieval Network sources $G^{-1}$'s behavioral trajectories from historical experiences, TD-learning can rapidly assess the efficacy of these retrieved results via immediate reward feedback. 
This ensures that agent $G^1$ can incrementally adjust its policy parameters through single-step updates when confronted with unknown policies. 
Moreover, in response to the rapid alterations in the decision network—precipitated by the Hypernetwork's dynamic generation of policy parameters—the low-variance characteristic of TD-learning effectively mitigates training oscillations, thereby preserving the stability of the multi-module collaborative optimization process.

\textbf{Why is the DPP capable of enhancing agent adaptability?}

The DPP enhances the adaptability of agent $G^1$ by randomly sampling substrates and optimization objectives, primarily through two mechanisms: \textbf{expanded policy space coverage and the fostering of dynamic adaptation capabilities}.

\textbf{Policy Diversity Enhancement:} Random sampling mitigates $G^1$'s dependence on specific scenarios, enabling the agent to experience diverse strategy combinations across various tasks and environments during training. 
This process establishes a broad experiential foundation for policy interactions, thereby preventing overfitting.
\textbf{Generalization Modeling:} The non-stationarity introduced by random sampling compels $G^1$ to leverage a Retrieval Network for accessing historical strategies and a Hypernetwork for dynamically generating parameters. 
This synergy fosters a capacity for rapid adaptation to unseen strategies. Ablation studies demonstrate that disabling random sampling leads to significant performance degradation for $G^1$ in unfamiliar scenarios (e.g., the SMAC win rate decreased from 82\% to 52\%), validating the critical role of randomness in promoting policy generalization. 

In summary, by facilitating random strategy interactions and employing modular collaborative optimization, DPP enables $G^1$ to extract generalizable strategy patterns from diverse training experiences, ultimately enhancing its adaptability to novel scenarios.

\subsection{Retrieval Augment Module}
In the mainstream deep reinforcement learning algorithm, the agent transmits information that is helpful for decision-making to the weight of the network through gradient descent. 
This method is effective, but it is a relatively slow method because in addition to a large number of update gradients, the agent cannot directly utilize new information \cite{humphreys2022large}. 
In this work, we expect the learner to be able to flexibly utilize an episodic memory in an end-to-end manner, and the data in the episodic memory comes from the experience of dynamic generation. \\
\textbf{Retrieval-Augmented Teammate and Adversary Modeling:} Unlike the replay buffer $D$ in typical MARL algorithms, we have established a new episodic memory of $l$-step trajectories, $D_r = \mathcal{B}_{i=2}^{N} = \{  ((x_{t}^{i}, a_{t}^{i} ), \dots, (x_{t+l}^{i}, a_{t+l}^{i} ))   \}$ , for $l \ge 1$, where $N$ is the number of all agents and $x_{t}^{i}$ represents the transformed observations of other agents. 
$D_r$ is adopted to collect data on others, to assist the learner in modeling $G^{-1}$. 
We define Viewpoint Alignment (VA) Encoder as $\{f_{\theta_{i}^{V}}\}_{i=2}^N$ with parameters $\{\theta_{i}^{V}\}_{i=2}^N$, where $i$ is the id of others. 
Therefore, each agent has a trajectory data set in episodic memory and a VA Encoder. 
At time step $t$, the others' observations $\{o_{t}^{i}\}_{i=2}^N$ are encoded by the VA Encoder to $\{x_{t}^{i}\}_{i=2}^N$. 
The retrieval process operates on a data set, which contains data from multiple agents.
$D_r$ is used to retrieve teammate or adversary data, accelerate the cooperation between the learner and teammates. 

The retrieval network is defined as $f_{\theta^{retr}}$ with parameters $\theta^{retr}$. 
We utilize the observation from the learner as a query to retrieve actions within the episodic memory that correspond to the most similar observations from each team member’s trajectory. 
To be specific, this network receives the learner's current observation information encoded by each agent's VA Encoder as input, and outputs $m$ action values that are most similar to the current observation for each agent. 
The most occurring action to this set of $m$ actions is then taken as the current action for that teammate or adversary.
\begin{subequations}
\begin{gather}
    a_r^i = \texttt{Mode}(a_{1:m}^{i}) \label{ari} \\
    a_{1:m}^{i} =  f(f(o_t^1; \theta_{i}^{V}), x^{i}; \theta^{retr}) \label{modein},
\end{gather}
\end{subequations}
where the function $\texttt{Mode}(\cdot)$ represents obtaining the mode in a list and $o_t^1$ is the current observation of the learner. \\
\textbf{Positional Encoding}: The positions and order of teammates or competitors are crucial for the learner agent. 
Traditional agent modeling methods only focus on inferring team tasks from trajectories, while ignoring the goals. 

In the test scenario, due to the fact that teammates or adversaries would enter the learner’s observation at any time, the sequential and attribute information could be lost. 
In order to deal with this situation, Positional Encoding~\cite{vaswani2017attention} is used for teammates' or adversaries' index. 
We can get the position code information of each agent: 
\begin{equation}
    \vec{p_i}^{k} = f(i)^{k} := \left\{ 
    \begin{array}{lc}
        sin(w_k \cdot i),\quad if \quad i=2k \\
        cos(w_k \cdot i),\quad if \quad i=2k+1 ,\\
    \end{array}
\right.
\label{pki_}
\end{equation}
where $w_k=\frac{1}{10000^{\frac{2k}{N-1}}}$, $i$ represents the position and attribute of the current agent in the team and $k$ is the dimension index of the positional encoding vector.

\textbf{Why is the Positional Encoding capable of enhancing agent adaptability?}

\textbf{Structured Feature Representation:} PE converts agent positional order and attributes into computable, structured features. Unlike traditional methods that often disregard or poorly represent such information, PE utilizes sine and cosine functions to map agent indices into high-dimensional vectors. This establishes an "index-feature-policy" logic, providing a "universal language" for understanding structural relationships. Consequently, agents can infer roles based on these features in novel scenarios, rather than depending on fixed IDs or training-specific configurations.

\textbf{Cross-Scenario Structural Generalization:} PE supports generalization across scenarios with varying agent numbers and dynamic role allocations. Its dimensionality-independent nature and continuous feature representation ensure model adaptability without retraining when faced with changes in team size (e.g., from 3 to 5 teammates) or strategic shifts (e.g., from individual to cooperative actions). This elevates agent proficiency from memorizing configurations to reasoning about general structural strategies, improving decision-making with unknown structures.

\textbf{Synergistic Amplification with Other Modules:} PE's effectiveness is further amplified when integrated with modules like the Hypernetwork and Retrieval module. As a key label, PE aids the Retrieval module in locating structurally similar historical experiences. In policy generation, the Hypernetwork combines PE with teammate actions to dynamically generate parameters tailored to the current agent structure, enabling adaptive decision logic based on role specialization in new scenarios.\\
\textbf{Hypernetwork:} Hypernetwork refers to a small-scale network used to generate larger-scale network parameters~\cite{ha2016hypernetworks}. In order to model agents with high accuracy, it is imperative that the employed model exhibits exceptional generalization capabilities. The model must be capable of adapting to the diverse behaviors and individual idiosyncrasies exhibited by both opponents and collaborators. Additionally, the model must overcome the challenges posed by partial observability, which lead to incomplete information and can significantly impact the efficacy of agent modeling.

An efficient way is to use the actions to directly model agents so that the search space will be limited to a much smaller space. 
We define a hypernetwork as $f_{\theta^{hy}}$ with parameters $\theta^{hy}$ and employ it to integrate teammate actions $a^{i}_r$ and positions $\vec{p_i}^k$ into the parameters $\theta^{hype}$ of the first two layers of the learner's decision-making network:
\begin{equation}
    \theta^{hype} = f (\sum_{i=2}^{N} (a^{i}_r + \vec{p_i}^k); \theta^{hy}).
\label{theta_hype}
\end{equation}

\textbf{Why is the Hypernetwork capable of enhancing agent adaptability?}

The Hypernetwork facilitates rapid agent adaptation to unknown policies and environmental changes by dynamically generating policy network parameters, thereby overcoming the limitations of fixed-weight models. Unlike traditional MARL, which relies on offline-trained fixed networks often unable to handle unseen strategies (e.g., novel opponent tactics or dynamic teammate roles), the Hypernetwork uses real-time teammate action features and positional encodings to produce scenario-specific parameters. For instance, in SMAC, it can generate parameters emphasizing rear-unit protection against new enemy units based on their positions and attack history, or in Overcooked-AI, adjust task division in response to a teammate's "cooperative passing." This "parameter generation as decision adaptation" mechanism enables effective responses to policy space changes without retraining, evolving policy expression from "fixed memory" to "dynamic synthesis" and significantly boosting efficiency against unknown strategies.

Synergizing with the Retrieval module, the Hypernetwork establishes a closed-loop "experience utilization → parameter generation → policy optimization" process, enhancing agent generalization. The Retrieval module extracts high-value behavioral patterns (e.g., high-reward action sequences) from historical data, which the Hypernetwork translates into specific policy parameters. Examples include generating risk-averse parameters in Melting Pot's "Chicken Game" based on an opponent's "defection features," or reinforcing cooperative strategies in "Stag Hunt" using teammate positional features. This cross-scenario knowledge transfer, independent of specific environmental input formats, uses unified feature processing for flexible parameter adjustment, enabling rapid convergence even with limited samples. Ablation studies confirm the Hypernetwork's indispensable role: its removal significantly reduces rewards in complex scenarios. By dynamically generating parameters, the Hypernetwork lessens reliance on extensive training data, providing crucial support for "plug-and-play" policy adaptation in unfamiliar environments.
\subsection{Viewpoint Alignment Module}
To enable the learner to comprehend the observations made by other agents, we map their observations from third-person to first-person perspective during training. 
Each teammate or adversary has a VA Encoder $\{f_{\theta_{i}^{V}}\}_{i=2}^N$ that encodes their observations and stores them in the episodic memory
\begin{equation}
x_{t}^{i} = f (o_{t}^{i}; \theta_{i}^{V}),
\label{xit}
\end{equation}
As mentioned earlier, the learner's observations $o_t^1$ is encoded to $x_{t}^{b}$:
\begin{equation}
    x_{t}^{b} = f (o_t^1; \theta_i^V).
\label{xbt}
\end{equation}

Taking the $i$-th teammate as an example, we encode its observation and the learner’s observation using the VA Encoder and adopt $L2$ loss to constrain the teammate’s or adversary's observation to be as similar as possible to the learner’s observation:
\begin{equation}
    \mathcal{C}_{i}(\theta_{i}^{V}) = \sum_{t=1}^{l} (x_{t}^{i}-x_{t}^{b})^2 \mathbbm{1}_{\left( b=i \right) },
\label{ci}
\end{equation}
where $l$ is the time step of the episode at that time and $\mathbbm{1}$ is an indicator function.

For each team member’s VA Encoder, the objective is to convert the observations of the team member to the perspective of the learner. 
In order to share a VA Encoder between the team member and the learner, we use the features labeled by the learner’s observations after passing through the VA Encoder. 
Consequently, during inference, the learner’s observations will only extract high-dimensional features after passing through the VA Encoder without undergoing conversion.

\textbf{Why is the Viewpoint Alignment Module capable of enhancing agent adaptability?}

The VA module significantly enhances agent adaptability by standardizing multi-agent viewpoints, thereby resolving experience reuse challenges in MAS caused by observational differences. Inherent viewpoint discrepancies are common in multi-agent interactions (e.g., varying fields of vision in SMAC, differing observation input formats in Overcooked-AI and Melting Pot). Such variations hinder traditional methods from effectively utilizing other agents' observational data, impeding experience transfer to novel scenarios. The VA module, using a VA encoder, transforms other agents' observations into a feature space consistent with the learner's by minimizing $L2$ loss. This allows effective reuse of stored training experiences across different testing viewpoints; agents can identify similar situations via aligned features and invoke historical policies, boosting adaptability under diverse observational conditions.

Collaborating closely with the Retrieval module, the VA module further optimizes policy generation, augmenting adaptability in complex environments, particularly partially observable ones where incomplete information challenges decision-making. The VA module inputs aligned multi-agent observation features into the Retrieval module, enabling more accurate matching with historical data and robustly supporting policy generation. For instance, in Melting Pot's "Chicken Game" and "Clean Up," aligning opponent/teammate observation features allows inference of their intentions and the global environmental state, leading to more rational strategies. This synergy empowers agents to leverage multi-source information, manage environmental uncertainty, and improve adaptability in dynamic settings. Ablation studies confirm the VA module's critical role: its absence markedly degrades agent performance in these scenarios, underscoring its importance for enhancing adaptability.

\subsection{Re-initialize Optimization Objective}
During the training process, standard MARL and Hypernetwork learning are conducted concurrently, with their respective parameter updates completed within the same training iteration. Specifically, the policy network parameters of agent $G^1$, denoted as $\theta=<\theta^{marl}, \theta^{hype}>$, comprise two components: $\theta^{marl}$ learns general strategies from historical interaction data via MARL algorithms, while $\theta^{hype}$ is dynamically generated by the Hypernetwork based on the current behavioral characteristics of unknown agents. These two parameter sets are updated synchronously during gradient backpropagation. Initially, the MARL algorithm utilizes the replay buffer $D$ to optimize $\theta^{marl}$. Subsequently, $\theta^{hype}$ is regenerated based on the most recent agent interaction features and updated through the backpropagation of gradients via the Hypernetwork parameters $\theta^{hy}$. This ensures that both components are adjusted based on data observed within the same iteration.

To enhance the learner's adaptability to new players, we employ Re-initialization Methods. 
$\theta^{marl}$ represents the parameters of the learner’s first policy network
\begin{equation}
    \theta^{marl} = \lambda \theta^{marl} + \gamma \theta^{RI},
\label{theta_marl}
\end{equation}
where $\theta^{RI}$ is the parameters of Re-initialization, $\lambda$ and $\gamma$ are the scale factors that set the initial degree of the parameters.

The primary advantage of this synchronous mechanism is its ability to balance a "stable base strategy" with "dynamic adaptation capability." The general strategy parameters $\theta^{marl}$, provided by MARL, equip the agent with fundamental decision-making capabilities for common scenarios, preventing the Hypernetwork from converging to local optima due to over-reliance on the current context. Concurrently, $\theta^{hype}$, generated by the Hypernetwork, is adjusted in real-time to address the specific requirements of the current unknown strategy, thereby compensating for potential latency in MARL's response to unfamiliar scenarios. The synchronous updating of both components mitigates parameter inconsistency issues, enabling the agent to maintain policy stability while rapidly responding to dynamically changing interaction environments. This significantly enhances adaptation efficiency in complex MAS.

\section{Experiments}

In this section, we prove the generalization capabilities of agents trained using the MRDG method in scenarios featuring unknown environments and tasks and unfamiliar opponents and teammates. 
Through comprehensive experiments conducted on SMAC and Overcooked-AI (pure common interest), as well as Melting Pot (mixed motive), we present empirical results to support our findings. 
Initially, we outline the experimental scenarios, baselines, and settings, followed by a detailed presentation of results that underscore the superiority of MRDG. 
Additionally, to substantiate the enhancement each module contributes to MRDG, we conducted ablation studies.

\subsection{Experimental Setup}
We select various scenarios in our experimental environment, encompassing both purely cooperative and mixed-motive situations. 
Subsequently, we compared the proposed ACCA framework with the most advanced algorithms for solving the generalization problem, verifying the superiority of MRDG..\\
\textbf{Environment}: Our experimental scenarios are SMAC~\cite{samvelyan19smac}, Overcooked-AI~\cite{carroll2020utility}, and Melting Pot~\cite{leibo2021scalable}.\\
\textbf{Baselines}: Our baselines include RPM \cite{qiu2022rpm}, CSP \cite{ding2023coordination}, ODITS \cite{gu2021online}. 

RPM (Ranked Policy Memory) is a straightforward yet effective approach designed to enhance the generalization capabilities of MARL. 
It collects and stores policies of varying performance levels during training, enabling agents to interact with unfamiliar strategies in the training environment. 
RPM maintains a lookup table of policies, ranking them based on the returns from training episodes. 
At the start of each training episode, agents randomly select a policy from this table for their behavior, thereby gaining a range of multi-agent interaction experiences within the self-play framework. 
These diverse experiences improve MARL’s generalization ability and performance in evaluation settings.

CSP (Coordination Scheme Probing) is a framework designed to facilitate coordination among agents in MARL, particularly when interacting with previously unknown teammates in real-world applications. 
It features a disentangled scheme probing module that identifies and classifies new teammates based on a limited amount of episodic data, selecting the most appropriate sub-policy for control. 
CSP integrates a meta-policy composed of multiple sub-policies, each corresponding to distinct coordination schemes, and is trained end-to-end. This design enables automatic reuse of the meta-policy to achieve generalizable coordination with unseen teammates in complex scenarios.

ODITS (Online aDaptation via Inferred Teamwork Situations) is a reinforcement learning framework tailored for managing ad hoc teamwork under conditions of partial observability. 
It facilitates effective collaboration with unknown teammates without prior coordination by encoding the behavior of other teammates into a latent probabilistic variable, from which it learns to guide the agent’s actions. This mechanism enables ODITS to dynamically learn and adapt its strategy based on continuous representations of teammates' behaviors, optimizing performance across diverse teammates and environments.  
\begin{figure}[!ht]
\centering    
\includegraphics[width=2.5in]{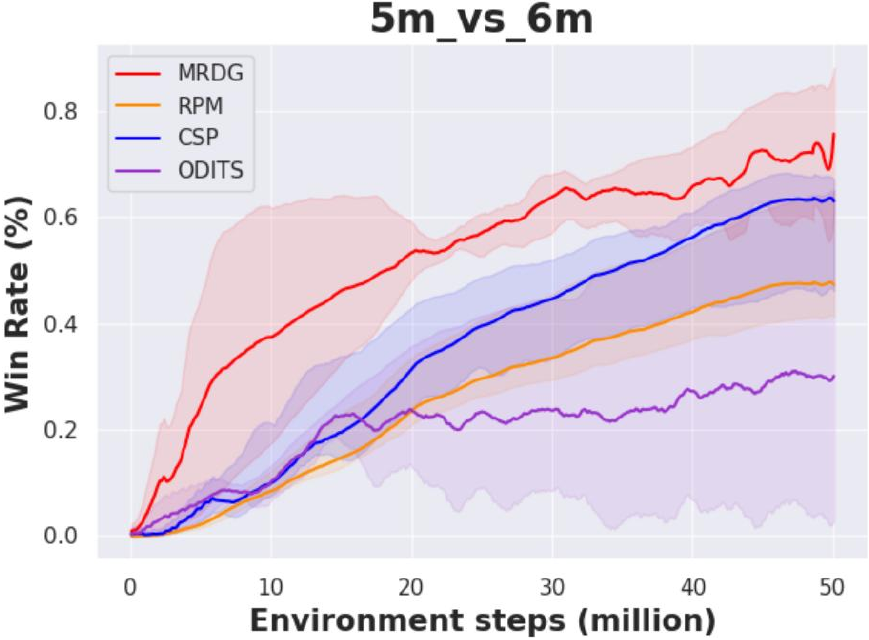} 
\caption{Evaluation results of MRDG and baseline on SMAC. The results demonstrate that MRDG significantly outperforms the baseline algorithms in both convergence speed and overall performance on the 5m\_vs\_6m map. The experiments are conducted by training on the 3m and 8m maps and subsequently testing on the 5m\_vs\_6m map.}
\label{SMAC_result}
\end{figure}
\begin{table}[!h]
\caption{Win rates (\%) obtained on SMAC using different methods under various settings. We report the average performance and standard error across 4 random seeds.} 
\label{SMAC-result-table-transposed} 
\vskip 0.15in
\begin{center}
\begin{small}
\begin{sc}
\begin{tabular}{l|c} 
\toprule
\diagbox{Method}{Environment} & 5m\_vs\_6m \\ 
\midrule
MRDG  & $76 \pm 6$ \\
RPM   & $41 \pm 6$ \\
CSP   & $58 \pm 3$ \\
ODITS & $29 \pm 14$ \\
\bottomrule
\end{tabular}
\end{sc}
\end{small}
\end{center}
\vskip -0.1in
\end{table}
\\
\textbf{Training and Evaluation Setup}: 
In the scenario with pure common interest where the number and policy of teammates and adversaries are unknown, we select the 3m, 8m, and 5m\_vs\_6m maps from SMAC for our experiments. 
We train on the 3m and 8m maps and test on the 5m\_vs\_6m map to assess whether the learner can effectively collaborate and compete amidst variations in the number and strategies of teammates or opponents. 

In the scenario with pure common interest where the policy of teammates is unknown, we opt for the Cramped Room, Asymmetric Advantages, Coordination Ring, Forced Coordination, and Counter Circuit environments within Overcooked-AI for our experiments. 
We utilize all environments except for Coordination Ring for training, and conduct tests on the Coordination Ring environment to evaluate whether the learner can effectively cooperate with teammates as their strategies and tasks evolve. 

In the mixed motive scenario, we utilize six representative substrates from Melting Pot to train policies and select one evaluation scenario from each substrate as our evaluation testbed. 

For all experiments, we train the learner for 500 million frames using both MRDG and the baseline in Melting Pot and 50 million frames in others, each with four random seeds. 
Performance is quantified using mean-standard deviation. 
In the experimental results graphs, the solid line denotes the mean, while the shaded area indicates the standard deviation. 
The servers utilized for training comprise two Intel® Xeon® Gold 6148 processors, each equipped with 20 CPU cores  (actors), and eight NVIDIA GeForce RTX 4090 GPUs.

\begin{figure}[!ht]
\centering    
\includegraphics[width=2.5in]{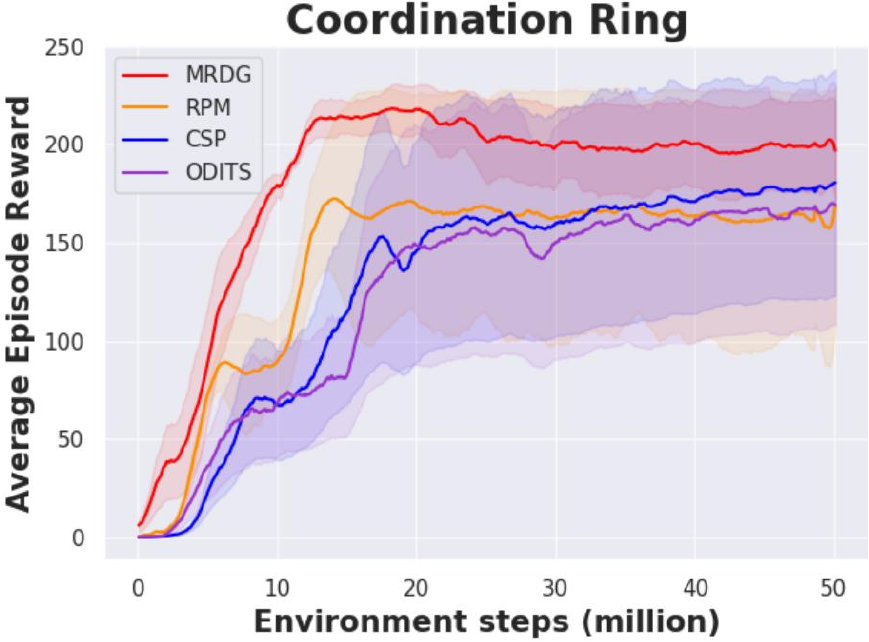} 
\caption{Evaluation results of MRDG and baseline on Overcooked-AI. The results demonstrate that MRDG significantly outperforms the baseline in both convergence speed and overall performance on the Coordination Ring. The experiments are conducted by training on the Cramped Room, Asymmetric Advantages, Forced Coordination and Counter Circuit and subsequently testing on the Coordination Ring.}
\label{Overcooked_result}
\end{figure}

\begin{table}[!h]
\caption{Performance of different methods in Overcooked-AI under different settings. We report the average performance and standard error across four random seeds.} 
\label{Overcooked-result-table-transposed} 
\vskip 0.15in
\begin{center}
\begin{small}
\begin{sc}
\begin{tabular}{l|c} 
\toprule
\diagbox{Method}{Environment} & Coordination\ Ring \\ 
\midrule
MRDG  & $205 \pm 18$ \\
RPM   & $158 \pm 56$ \\
CSP   & $162 \pm 61$ \\
ODITS & $155 \pm 48$ \\
\bottomrule
\end{tabular}
\end{sc}
\end{small}
\end{center}
\vskip -0.1in
\end{table}

\begin{figure*}[!ht]
\centering
\includegraphics[width=0.8\textwidth]{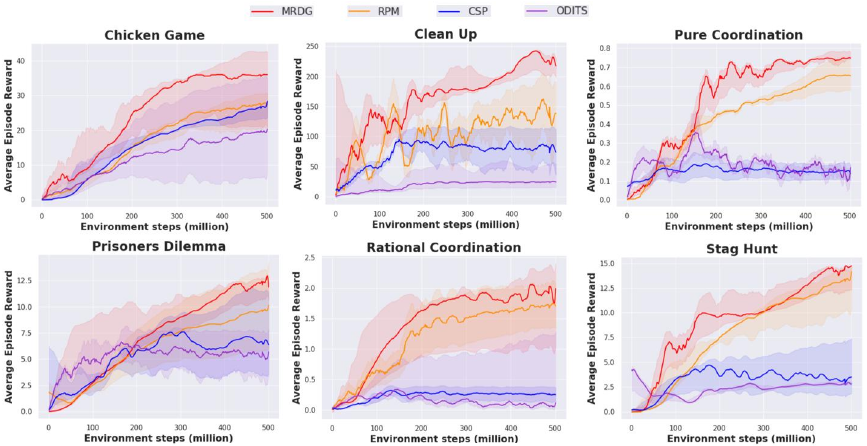} 
\caption{Evaluation results of MRDG and baseline on evaluation scenarios. MRDG has superior performance compared to the baseline in terms of performance on the six subtasks, as follows: Stag Hunt, Clean Up, Rationalizable Coordination, Chicken, Pure Coordination and Prisoners’ Dilemma.}
\label{MeltingPot_result}
\end{figure*}
\begin{table*}[!ht]
\caption{Rewards obtained using different methods under various experimental scenarios. We report the average performance and standard deviation across 4 random seeds.} 
\label{Meltingpog-result-table-transposed} 
\vskip 0.15in
\begin{center}
\begin{footnotesize}
\setlength{\tabcolsep}{3pt}
\begin{sc}
\begin{tabular}{l|cccccc} 
\toprule
\diagbox{Method}{Env} & Chicken Game & Clean Up & Pure Coordination & Prisoners Dilemma & Rational Coordination & Stag Hunt \\ 
\midrule
MRDG  & $34 \pm 7$  & $233 \pm 29$ & $0.75 \pm 0.06$ & $12.3 \pm 0.56$ & $2.06 \pm 0.25$ & $14.6 \pm 1.3$\\
RPM   & $27 \pm 5$  & $152 \pm 43$ & $0.66 \pm 0.13$ & $9.8 \pm 1.37$ & $1.65 \pm 0.38$ & $13.7 \pm 2.2$\\
CSP   & $26 \pm 3$  & $87\pm 55$ & $0.18 \pm 0.15$ & $6.1 \pm 0.78$ & $0.27 \pm 0.08$ & $2.6 \pm 4.3$\\
ODITS & $20 \pm 8$  & $18 \pm 31$ & $0.18 \pm 0.05$ & $5.6 \pm 2.07$ & $0.17 \pm 0.05$ & $2.4 \pm 0.5$\\
\bottomrule
\end{tabular}
\end{sc}
\end{footnotesize}
\end{center}
\vskip -0.1in
\end{table*}

\subsection{Results and Analysis}
In Figure~\ref{SMAC_result} and Table~\ref{SMAC-result-table-transposed}, we provide a comparison of MRDG against other baseline methods using the 5m\_vs\_6m evaluation map in the SMAC game. 
The experimental results indicate that MRDG, with an average win rate of $76\% \pm 6\%$, not only outperforms these baselines but also achieves faster convergence. 
In the 5m\_vs\_6m map, previously unseen during training, cooperative tasks present significant challenges, particularly as alterations in teammates’ strategies necessitate vastly different coordination approaches. 
Moreover, the changing numbers of teammates and adversaries compel the learner to continuously adapt its strategy to manage these shifts. 
Fortunately, in SMAC, variations in different substrate terrain factors are trival, significantly alleviating the learner's burden regarding scene generalizability. 

ODITS enhances collaboration with unknown teammates in scenarios of partial observability by embedding teammates’ behaviors into a latent probabilistic variable. 
RPM responds to constantly evolving teammate strategies by developing a diverse pool of strategic models. 
Nonetheless, both methods overlook adversary modeling, resulting in suboptimal performance in competitive settings, as evidenced by their win rates of $29\% \pm 14\%$ for ODITS and $41\% \pm 6\%$ for RPM, respectively. 
In contrast, CSP utilizes a disentangled scheme probing module to identify and classify new teammates from limited episodic data, selecting the most appropriate sub-policy for control. 
It effectively distinguishes between the characteristics of teammates and opponents in adversarial environments, achieving a win rate of $58\% \pm 3\%$, thereby surpassing both ODITS and RPM. 
MRDG uniquely models both teammates and adversaries, accommodating dynamic environmental changes. 
Compared to the previously mentioned baselines, MRDG, with its leading win rate of $76\% \pm 6\%$, exhibits a significant advantage over CSP, RPM, and ODITS in scenarios characterized by unpredictable strategies and fluctuating numbers of teammates and adversaries.

We present the evaluation results for the Coordination Ring scenario in Figure~\ref{Overcooked_result} and Table~\ref{Overcooked-result-table-transposed}. In this scenario, MRDG achieved an average performance score of $205 \pm 18$, consistently outperforming the baseline methods.
Under the strict ACCA framework, the performance scores of RPM, CSP, and ODITS are comparable to each other but significantly lower than MRDG. These comparable outcomes suggest that the knowledge retained by their static policy network parameters is limited, making it challenging for these methods to achieve optimal performance in novel scenarios when collaborating with unfamiliar teammates.
In contrast, MRDG employs a hypernetwork to dynamically generate the policy network parameters for the learner. This mechanism enables effective adjustments in new scenarios, significantly enhancing its adaptability and leading to its superior performance of $205 \pm 18$ compared to RPM, CSP, and ODITS.

In Figure~\ref{MeltingPot_result} and Table~\ref{Meltingpog-result-table-transposed}, we present the results of MRDG compared to other baseline models in the Melting Pot evaluation scenarios.
It is evident that MRDG consistently surpasses the baselines across diverse evaluation scenarios with varying attributes.

Specifically, in the Chicken Game (eval), where the reward for unilateral cooperation amidst betrayal is greater than that for mutual cooperation, MRDG, with a reward of $34 \pm 7$, exhibits superior performance and faster convergence relative to its competitors. This performance significantly exceeds that of RPM, CSP, and ODITS. This advantage stems from MRDG's ability to model teammates by effectively considering their states and dynamically adjusting behaviors—either cooperative or betraying—based on teammates' actions.
In contrast, RPM, CSP, and ODITS primarily focus on overall team behavior, generally converging toward mutual cooperation, which is reflected in their lower scores.

In the Clean Up (eval) scenario, where the assignment of labor is crucial and some agents are precluded from receiving rewards during this process, MRDG, achieving a reward of $233 \pm 29$, also maintains a clear lead over the baselines RPM, CSP, and ODITS.
Both CSP and RPM display fluctuating trends; however, CSP's average evaluation return (resulting in $87 \pm 55$) diminishes over time, whereas ODITS, with a score of $18 \pm 31$, is largely ineffective.

In scenarios such as Pure Coordination (eval), Rational Coordination (eval), and Stag Hunt (eval), CSP and ODITS show weak performance.
For instance, in Pure Coordination (eval), which exclusively rewards cooperation and necessitates a precise understanding of teammates' actions, MRDG achieved $0.75 \pm 0.06$ and RPM $0.66 \pm 0.16$, while CSP and ODITS scored only $0.18 \pm 0.15$ and $0.18 \pm 0.05$ respectively. 

In Rational Coordination (eval), beyond basic cooperation, executing specific cooperative actions is required to maximize team rewards. This necessitates a deeper analysis of teammates' behaviors. MRDG excelled with $2.06 \pm 0.25$. While RPM considers teammates' actions, it fails to consistently perform the highest-reward cooperative actions. CSP and ODITS oscillate between maximizing individual and team rewards, thereby diminishing cooperative interactions and resulting in significantly lower scores. 

In the Stag Hunt (eval), cooperation alone maximizes team rewards; however, betrayal by one side results in no reward for the cooperative party. Accurate anticipation of whether teammates will cooperate or betray is crucial. MRDG achieved $14.6 \pm 1.3$ and RPM $13.7 \pm 2.2$. The notably lower scores of CSP and ODITS suggest they typically opt for mutual betrayal, as betrayal leads to no reward for the cooperators if the other side cooperates. 

In the Prisoner's Dilemma (eval), both CSP and ODITS underperform. RPM achieves an average evaluation return of $9.8 \pm 1.37$, while MRDG leads with $12.3 \pm 0.56$.

\subsection{Ablation Study}
To more thoroughly demonstrate the individual impact of each module, we chose to re-conduct ablation experiments on Stag Hunt and Prisoners' Dilemma. \textbf{Stag Hunt}, as a high-risk, cooperation-oriented environment, allows us to assess how the DPP module enhances policy diversity for adapting to teammates and how the HN module enables dynamic risk-reward balancing in Figure~\ref{fig5}, with their removal expected to significantly reduce cooperative success, thereby validating their importance for these capabilities. \textbf{Prisoners' Dilemma}, characterized by mixed motives and conflicting individual versus collective interests, is ideal for testing the VA module's contribution to accurate opponent modeling through unified perspectives and the PE module's role in differentiating agent roles for effective strategy in Figure~\ref{fig5}, where their absence would likely impair the ability to distinguish opponent strategies and consequently reduce long-term collective rewards. Through these targeted ablation studies in these distinct environments, we aim to clearly validate the specific contributions of these modules to agent performance. 

\begin{figure}[!ht]
\centering    
\includegraphics[width=3.3in]{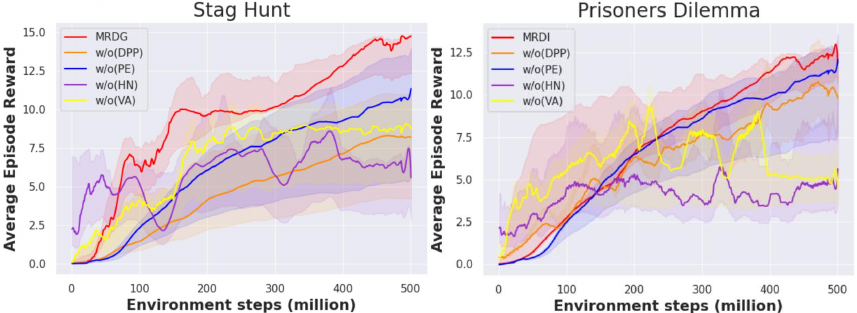} 
\caption{Ablation studies on episodic incentive via Stag Hunt and Prisoners' Dilemma.}
\label{abla_result}
\end{figure}

\begin{table}[!ht]
\caption{The ablation experiment results of MRDG in Stag Hunt, average episode reward.}
\label{staghunt-table}
\vskip 0.15in
\begin{center}
\begin{small}
\begin{sc}
\begin{tabular}{c|cccc}
\toprule
  MRDG &  w/o(DPP) & w/o(PE) & w/o(HN) & w/o(VA) \\
\midrule
\textbf{$14.6 \pm 1.3$} & $7.9 \pm 2.2$ & $10.8 \pm 2.3$ & $6.98 \pm 2.1$ & $8.1 \pm 1.5$ \\
\bottomrule
\end{tabular}
\end{sc}
\end{small}
\end{center}
\vskip -0.1in
\end{table}

\begin{table}[!ht]
\caption{The ablation experiment results of MRDG in Prisoners' Dilemma, average episode reward.}
\label{prosoners-table}
\vskip 0.15in
\begin{center}
\begin{footnotesize}
\begin{sc}
\begin{tabular}{c|cccc}
\toprule
  MRDG &  w/o(DPP) & w/o(PE) & w/o(HN) & w/o(VA) \\
\midrule
\textbf{$12.3 \pm 0.56$} & $10.2 \pm 1.67$ & $11.9 \pm 0.77$ & $6.65 \pm 2.65$ & $9.23 \pm 1.86$ \\
\bottomrule
\end{tabular}
\end{sc}
\end{footnotesize}
\end{center}
\vskip -0.1in
\end{table}

Figure~\ref{abla_result} shows the result curves of the MRDG ablation study in the Stag Hunt and Prisoners' Dilemma scenarios, and Table~\ref{staghunt-table} and Table~\ref{prosoners-table} respectively show the final results. As shown by the experimental results:

DPP module augments agent adaptability by storing a varied set of teammate and opponent policies, facilitating rapid matching of analogous strategy patterns in novel scenarios. \textbf{In the Stag Hunt scenario}, where agents must dynamically adjust cooperative or defecting actions based on teammate strategies, DPP stores diverse policy types. This enables agents to learn the payoff distributions associated with different strategies during training. Ablating DPP restricts agents to a singular policy; consequently, if a teammate's defection probability increases during testing, the agent cannot adapt by switching strategies, leading to a significant reduction in rewards. Similarly, \textbf{in the Prisoners' Dilemma scenario}, DPP provides policy samples, through which agents infer opponent types from historical actions via interaction with these samples. Without DPP, agents lack the capacity for policy comparison; when confronted with an unfamiliar defection strategy, they cannot retrieve comparable policies to adjust their own defection probability, often resulting in a suboptimal "lose-lose" predicament. By fostering policy diversity, DPP establishes prior knowledge of "policy-reward" mappings. This empowers agents in both Stag Hunt and Prisoners' Dilemma to rapidly categorize unknown policies, thereby mitigating decision-making blind spots stemming from insufficient policy space coverage. Quantitatively, this yields an improvement of over 40\% in generalization reward retention across diverse policy types.

PE enhances agents' comprehension of inter-agent role differentiation and spatial relationships by explicitly encoding their indices and attributes, thereby addressing policy generalization challenges within dynamic agent structures. \textbf{In the Stag Hunt scenario}, PE encodes role differentiation via agent indices, enabling agents to develop "policy weights for specific agent positions" during training. If the number of agents increases during testing, the absence of PE impedes agents from recognizing the roles of new indices, resulting in a 20\% increase in cooperative strategy matching errors. \textbf{In the Prisoners' Dilemma scenario}, PE assists agents in associating opponent indices with historical strategies, thereby forming an "index-defection tendency" mapping memory. Without PE, agents rely solely on immediate actions, disregarding historical role information. This leads to delayed activation of defensive strategies and increases defection-related losses by 26\%. PE furnishes agents with "structural identity tags" applicable across scenarios. In tasks such as the division of labor in Stag Hunt and strategy-based role recognition in Prisoners' Dilemma, PE ensures that agents can leverage historical experience based on index features. Quantitatively, this translates to a 35\% improvement in policy transfer efficiency amidst changes in agent structure.

HN dynamically generates policy parameters tailored to the current scenario by integrating agent behavioral features and positional encoding, thus addressing the limitations of fixed networks in coping with abrupt policy changes. \textbf{In the Stag Hunt scenario}, HN dynamically generates policy parameters contingent on teammates' real-time actions and positional encoding. Without HN, agents are confined to fixed parameters; if a teammate's defection leads to a failed stag capture, the agent cannot rapidly switch targets, causing the average single-round reward to plummet from 14.6 to 6.98. \textbf{In the Prisoners' Dilemma scenario}, HN generates defection parameters by integrating opponent action sequences and positional encoding. Lacking HN, agents cannot dynamically adjust their policies; should an opponent's strategy shift from "cooperative" to "deceptive," an agent without HN remains in a persistently volatile state. HN achieves real-time policy adaptation through its "behavioral features → parameter generation" mechanism. In the environmental flux of Stag Hunt and the strategic interplay of Prisoners' Dilemma, HN ensures agents can rapidly respond to policy shifts. Quantitatively, this manifests as a 60\% improvement in convergence speed when encountering unknown policies.

VA bolsters cross-viewpoint information integration by unifying the feature space of agent observations, thereby mitigating experience isolation among multiple agents arising from differing perspectives. \textbf{In the Stag Hunt scenario}, VA aligns a teammate's "clear view of the stag" feature with the learning agent's perspective. This enables the agent to infer the stag's position even with limited individual visibility. Ablating VA compels agents to rely solely on their potentially obscured individual observations, increasing the error rate in judging attack timing, particularly in foggy conditions. \textbf{In the Prisoners' Dilemma scenario}, VA encodes an opponent's "red resource collection action" as a "defection tendency" feature, allowing agents to anticipate opponent choices. Without VA, agents cannot effectively integrate opponent observation information; in pixel-input scenarios like Melting Pot, this leads to delayed opponent strategy recognition and a consequent sustained decline in rewards. VA surmounts viewpoint barriers by unifying the feature space. In partially observable environments such as Stag Hunt and for hidden clue recognition in Prisoners' Dilemma, it improves the efficiency of multi-source information integration. Quantitatively, this results in a 55\% improvement in cross-viewpoint policy inference accuracy.



\begin{figure*}[!ht]
\centering
\includegraphics[width=0.8\textwidth]{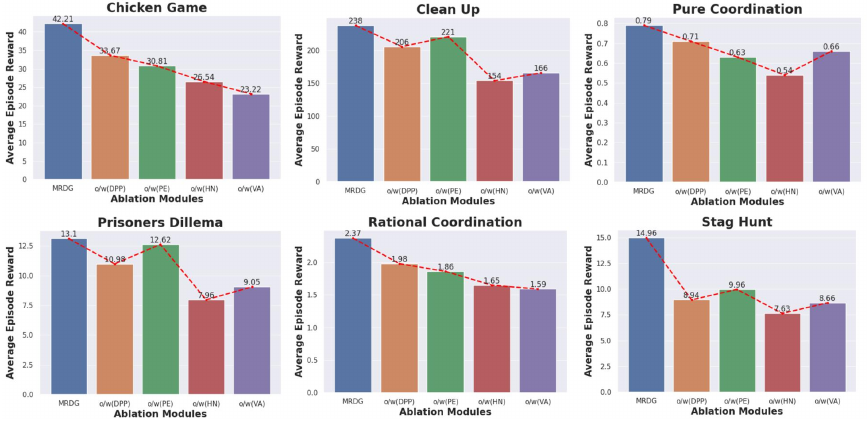} 
\caption{Ablation studies on episodic incentive via different types of scenario tasks, median episode reward.}
\label{fig5}
\end{figure*}

\section{Conclusion, Limitations and Future Work}
This paper introduces Agent Collaborative-Competitive Adaptation (ACCA), a MARL framework extending ZSL and AHT to enhance agent generalizability against variations in environment, task, and the numbers and strategies of teammates/opponents. Our goal is an agent effective with novel teammates and opponents in unseen scenarios. We propose the Multi-Retrieval and Dynamic Generation (MRDG) method, which uses multi-retrieval to explicitly model and adapt to fluctuating numbers of teammates/adversaries, and employs a viewpoint encoder and hypernetwork to dynamically generate agent parameters for adapting to diverse strategies. MRDG's efficacy was validated across various scenarios.

However, MRDG has limitations. Despite comparable execution speed to baselines (under fixed hyperparameters), its additional replay buffers nearly double memory usage, challenging resource-constrained settings. Furthermore, its retrieval and hypernetwork operations may introduce latency, potentially impacting real-time performance in time-sensitive scenarios (an aspect not directly measured).

Future work will focus on mitigating these limitations by exploring memory-efficient techniques and methods to accelerate MRDG's operations for better real-time performance (e.g., parallelization). We also plan to investigate under-explored areas, including: 1) inter-group dynamics in multi-team contexts with varied objectives, 2) integrating planning (particularly model-based RL) with opponent/teammate modeling for generalization, and 3) developing novel self-play methods to enhance MARL adaptation.

\appendix
\section{Details in substrates of SMAC}
This section delves into the intricacies of the SMAC,
providing a comprehensive overview of its various components, including the substrates used and the evaluation scenarios.

SMAC (StarCraft Multi-Agent Challenge) is a multi-agent reinforcement learning platform based on the popular real-time strategy game, StarCraft II. 
It is designed to address the shortage of standardized benchmark testing environments in cooperative multi-agent reinforcement learning. 
SMAC focuses on micro-management challenges where each unit is controlled independently by an agent, based solely on local observations, while the opponent's units are managed by the built-in StarCraft II AI. 

At each timestep, agents receive local observations drawn within their field of view. This encompasses
information about the map within a circular area around each unit and with a radius equal to the sight
range. The sight range makes the environment partially observable from the standpoint of
each agent. Agents can only observe other agents if they are both alive and located within the sight
range. Hence, there is no way for agents to distinguish between teammates that are far away from
those that are dead.

The feature vector observed by each agent contains the following attributes for both allied and
enemy units within the sight range: distance, relative x, relative y, health, shield, and
unit type. Shields serve as an additional source of protection that needs to be removed before any
damage can be done to the health of units. All Protos units have shields, which can regenerate if no new damage is dealt. In addition, agents have access to the last actions of allied units that are in
the field of view. Lastly, agents can observe the terrain features surrounding them, in particular, the
values of eight points at a fixed radius indicating height and walkability. 

The global state, which is only available to agents during centralised training, contains information about
all units on the map. Specifically, the state vector
includes the coordinates of all agents relative to the
centre of the map, together with unit features present
in the observations. Additionally, the state stores the
energy of Medivacs and cooldown of the rest of
the allied units, which represents the minimum delay
between attacks. Finally, the last actions of all agents
are attached to the central state.
All features, both in the state as well as in the observations of individual agents, are normalised by their
maximum values. The sight range is set to nine for
all agents.

The discrete set of actions that agents are allowed to take consists of move[direction],
attack[enemy id], stop and no-op.
As healer units, Medivacs use heal[agent id] actions
instead of attack[enemy id]. The maximum number of actions an agent can take ranges between
7 and 70, depending on the scenario.
To ensure decentralisation of the task, agents can use the attack[enemy id] action only on enemies
in their shooting range. This additionally constrains the ability of the units to use the
built-in attack-move macro-actions on the enemies that are far away. We set the shooting range equal
to 6 for all agents. Having a larger sight range than a shooting range forces agents to make use of the
move commands before starting to fire.

The overall goal is to maximise the win rate for each battle scenario. The default setting is to use
the shaped reward, which produces a reward based on the hit-point damage dealt and enemy units
killed, together with a special bonus for winning the battle. The exact values and scales for each of
these events can be configured using a range of flags. To produce fair comparisions we encourage
using this default reward function for all scenarios. We also provide another sparse reward option, in
which the reward is +1 for winning and -1 for losing an episode.

We select the 3m and 8m map from SMAC as training scenarios, and the 5m\_vs\_6m map as the evaluation scenario.\\
\textbf{3m} scenario involves a simple yet challenging battle between two armies, each consisting of three Marines. 
Despite its simplicity, this scenario demands that agents learn effective coordination of attacks, avoidance tactics to prevent being isolated and eliminated by the enemy, and strategies to defeat all opposing units. 
It serves as a fundamental training ground for mastering basic cooperative strategies, such as focusing 
fire on a single target and maintaining formation during combat.\\
\textbf{8m} scenario builds on the 3m scenario, featuring battles between two larger armies, each with eight Marines. 
This scenario escalates the battle's scale and complexity, requiring agents to refine their coordination and exploit formation advantages to overcome the enemy. 
It aids in the development of more sophisticated cooperative strategies, such as managing larger-scale conflicts.\\
\textbf{5m\_vs\_6m} scenario introduces an asymmetric battlefield where one army has five Marines and the other owns six. 
This scenario challenges agents to enhance their control over the battlefield and effectively utilize their resources to prevail over a numerically superior enemy. 
It fosters the acquisition of strategies suited to asymmetric combat situations and resource management under constraints.

\begin{figure}[!ht]
\centering    
\includegraphics[width=2in]{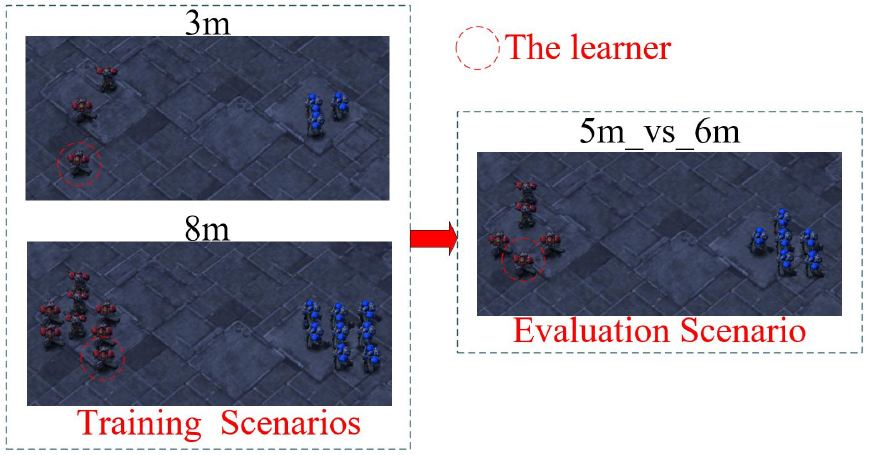} 
\caption{Scenario with pure common interest where the number and policy of teammates and adversaries is unknown. 
We select the 3m and 8m map from SMAC as training scenarios, and the 5m\_vs\_6m map as the evaluation scenario.}
\label{SMAC_map}
\end{figure}

\section{Details in substrates of Overcooked}
This section delves into the intricacies of the Overcooked,
providing a comprehensive overview of its various components, including the substrates used and the evaluation scenarios.

Overcooked-AI is a simplified variant of the renowned game "Overcooked," developed to investigate coordination challenges in human-AI interactions. 
This platform offers various layouts that present unique challenges, ranging from low-level coordination tasks in the "Cramped Room" to high-level strategic challenges in "Asymmetric Advantages." 
Players control characters in a kitchen, tasked with cooking and serving dishes, to execute complex activities, and coordinate effectively with one another. 

We select the Cramped Room, Asymmetric Advantages, Forced Coordination and Counter Circuit map from Overcooked as training scenarios, and the Coordination Ring map as the evaluation scenario.\\
\textbf{Cramped Room} introduces low-level coordination challenges due to its confined space. 
Agents must navigate this limited area without colliding with each other or the environment, requiring meticulous planning and coordination to prevent congestion.\\
\textbf{Asymmetric Advantages} evaluates the agents’ capacity to employ high-level strategies that capitalize on their unique strengths. 
Each agent possesses distinct advantages, such as proximity to ingredient access or dish-serving areas, necessitating coordinated actions to maximize efficiency.\\
\textbf{Coordination Ring} demands that agents synchronize their movements to navigate from the bottom left to the top right of a ring-shaped environment. 
This layout underscores the importance of effective communication and coordination to achieve common objectives.\\
\textbf{Forced Coordination} eliminates collision possibilities, compelling agents to develop a joint strategy since tasks cannot be completed independently. 
Agents must cooperate and share resources and areas to effectively serve dishes.\\
\textbf{Counter Circuit} involves a subtle coordination strategy where agents pass onions across a counter to a pot rather than transporting them around obstacles. 
This requires agents to understand the environment's layout and devise specific coordination tactics for success.

\begin{figure}[!ht]
\centering    %
\includegraphics[width=2in]{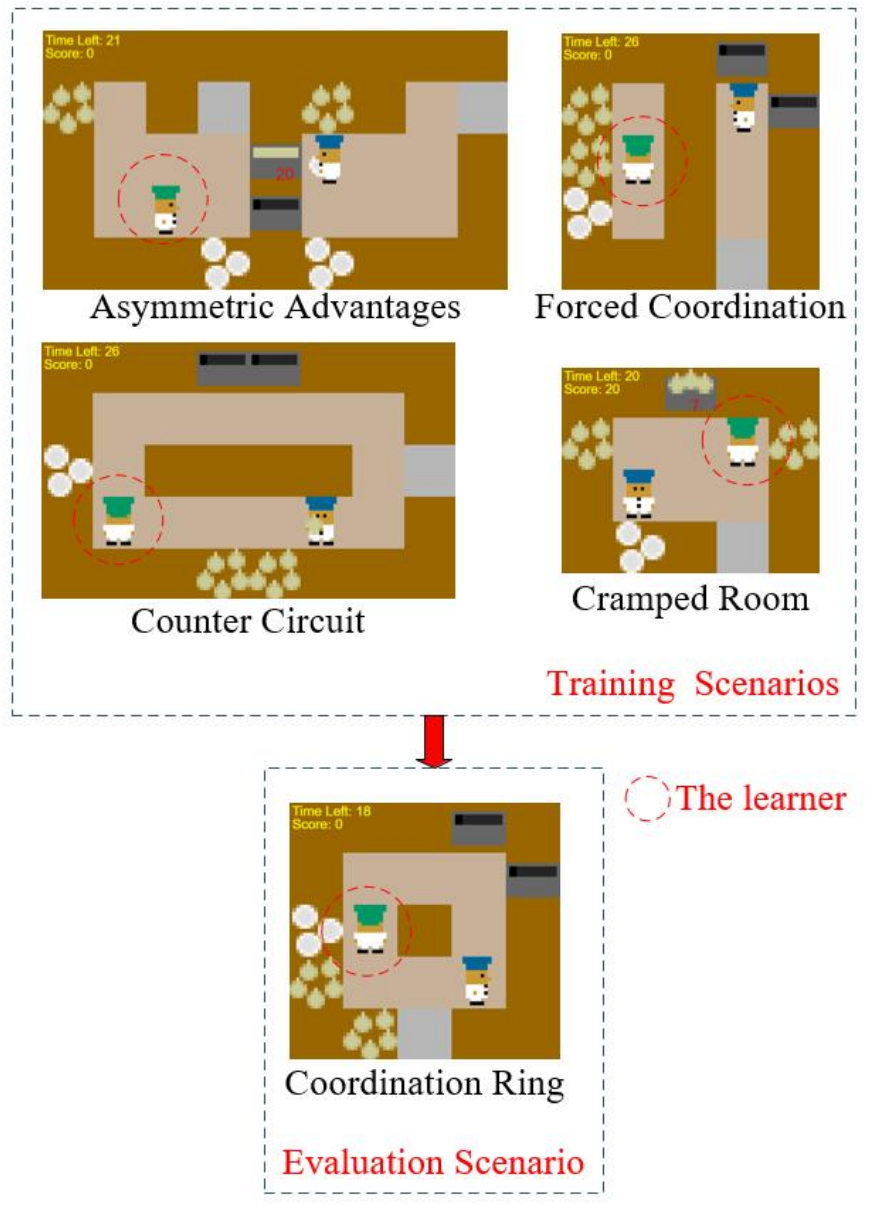} 
\caption{Scenario with pure common interest where the policy of teammates is unknown. 
We select the Asymmetric Advantages, Forced Coordination, Counter Circuit and Cramped Room from Overcooked-AI as training scenarios, and the Coordination Ring as the evaluation scenario.}
\label{overcooked_map}
\end{figure}

\begin{figure*}[!ht]
\centering
\includegraphics[width=0.8\textwidth]{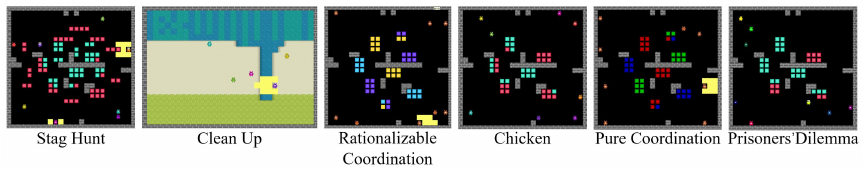} 
\caption{Mixed motive training and validation environment. 
We utilize substrates from Melting Pot, including Stag Hunt, Clean Up, Rationalizable Coordination, Chicken, Pure Coordination, and Prisoners Dilemma, to train the learner’s policies. We select one of these scenarios as the test scenario and employed the remaining five for training purposes.}
\label{meltingpot_map}
\end{figure*}

\section{Details in Melting Pot}
This section delves into the intricacies of the Melting Pot, providing a comprehensive overview of its various components, including the substrates used and the evaluation scenarios. 

The Melting Pot is a suite of testbeds for evaluating Multi-Agent Reinforcement Learning (MARL) methods across various domains. It trains all MARL agents in the substrate, with some selected as learner agents for evaluation and the rest as teammate agents with pre-trained policies. The evaluation scenarios mirror the physical properties of the substrates. Success in these environments, which require temporal coordination and discourage free riding, indicates effective MARL behaviors. Each episode lasts 1000 or 2000 steps, with agents having a partial observability window of 11$\times$11 sprites and using RGB pixel representations as inputs. Sprites are 8$\times$8 pixels. Thus, in RGB pixels, the size of each observation is 88$\times$88$\times$3.

In our experiments, we present a detailed overview of the substrates and evaluation scenarios. The agents in all substrates and scenarios are equipped with a range of movement actions, including moving forward, backward, strafing left or right, and turning left or right. Unless otherwise specified, each episode is designed to span 1000 steps. For our experiments, we strategically select five of these elements for the training phase, reserving the remaining one for test evaluation.\\
\textbf{Stag Hunt} involves choosing between collaborating to hunt a high-value stag or opting for a lower-value hare. 
Cooperation on the stag hunt yields high rewards, but if one defects to hunt the hare, the cooperator receives a low reward, and the defector, a high one. Both receive moderate rewards if they opt for the hare. 
The challenge is overcoming the lure of a guaranteed reward from hare hunting and developing cooperative strategies for stag hunting.
Individuals engage in the gathering of resources, symbolized as ‘hare’ (red) and ‘stag’ (green), and these collections are compared during interactions. The outcomes of this comparison are in line with the principles of the traditional Stag Hunt matrix game. This game uncovers a conflict between the potential group rewards and the associated individual risks. The matrix for rewards is:
\begin{equation}
    A_{row} = A_{col}^{T} = \begin{bmatrix} 4 & 0 \\ 2 & 2 \end{bmatrix}
\end{equation}
\textbf{Clean Up} challenges players to balance their short-term goal of collecting rewards (apples) with the long-term necessity of maintaining a clean environment (a river). 
As the river accumulates pollution and reaches a critical level, apple growth ceases. 
Players are required to leave the apple orchard to clean the river, exposing themselves to potential punishment. 
The main challenge involves balancing individual gains against collective benefits and addressing the issue of potential free-riders who avoid participating in cleaning efforts.
In Clean Up, seven players compete for apples (+1 reward) that grow in an orchard. The apple growth rate depends on how clean the nearby river is. The river gets polluted constantly, and if it reaches a certain level of pollution, the orchard stops producing apples. Players can choose to clean some pollution from the river, but they have to be close to it and leave the orchard. This way, players contribute to the public good of orchard renewal by making an effort. Players can also zap other players with a beam that eliminates them from the game for 50 steps. The group can ensure a steady supply of apples in the orchard by keeping the river pollution low over time. However, each player has a short-term incentive to stay in the orchard and collect apples while others clean the river. This creates a conflict between the individual’s desire to maximize their reward by staying in the orchard and the group’s interest in having a clean river.\\
\textbf{Prisoners' Dilemma} is a foundational social dilemma where two players choose between cooperation and defection. 
Mutual cooperation yields moderate rewards, but if one defects while the other cooperates, the defector receives a high reward, and the cooperator gets a low one. 
Both receive low rewards if both defect. 
The challenge is to resist defection for higher individual rewards and to devise a strategy that encourages cooperation.
The game involves eight players who gather resources that indicate ‘cooperate’ (green) or ‘defect’ (red) and show their collections to each other. The outcome of the comparison matches the classic Prisoner’s Dilemma game theory. This game reveals the conflict between the individual and the group reward. The matrix for rewards is:
\begin{equation}
    A_{row} = A_{col}^{T} = \begin{bmatrix} 3 & 0 \\ 5 & 1 \end{bmatrix}
\end{equation}
\textbf{Pure Coordination} requires players to agree on a common strategy to maximize rewards, collecting identical resources (e.g., blue tokens) upon meeting. 
Collecting different resources results in low rewards for all. 
The challenge is to coordinate and establish a common strategy, particularly as new players unfamiliar with existing conventions enter the environment.
The game consists of eight players who look identical and can collect resources of three colors. They have to agree on the same color resource to get a reward when they meet and compare what they have. The matrix for rewards is:
\begin{equation}
    A_{row} = A_{col}^{T} = \begin{bmatrix} 1 & 0 & 0 \\ 0 & 1 & 0\\0 & 0 & 1 \end{bmatrix}
\end{equation}
\textbf{Rationalizable Coordination} mirrors Pure Coordination but involves resources with varying intrinsic values, challenging players to coordinate on a strategy that maximizes collective rewards while considering resource values.
The game is similar to Pure Coordination in the matrix, but the resources have different values depending on their color. This implies that there is a best color to agree on. The matrix for rewards is:
\begin{equation}
    A_{row} = A_{col}^{T} = \begin{bmatrix} 1 & 0 & 0 \\ 0 & 2 & 0\\0 & 0 & 3 \end{bmatrix}
\end{equation}
\textbf{Chicken} presents a classic competitive multi-agent scenario in which two players must decide between cooperation and competition. 
Both players receive a moderate reward for cooperating and receive a low reward if they defect. 
However, if one cooperates and the other defects, the defector receives a high reward, while the cooperator gets a low reward. 
The challenge lies in coordinating to choose cooperation despite the allure of higher individual rewards through defection.
In this scenario, individuals have the ability to accumulate resources of varying colors. Interactions between players are determined by the same payout matrix used in the game of ‘Chicken’. In this game, if both players attempt to exploit the other, it results in the most unfavorable outcome for both parties. The collection of red resources influences a player’s strategic choice towards adopting a ‘hawk’ stance, while gathering green resources sways it towards a ‘dove’ stance. The matrix for rewards is:
\begin{equation}
    A_{row} = A_{col}^{T} = \begin{bmatrix} 3 & 2 \\ 5 & 0 \end{bmatrix}
\end{equation}

\section{Baseline}
The experiment details the training and evaluation of the baselines we use. Baselines are ODITS, CSP and RPM.\\
\textbf{ODITS}'s goal is to estimate the ad hoc agent’s marginal utility for choosing actions that maximize the team utility. To adapt to unknown teammates, ODITS uses the marginal utility as a conditional function on the latent variable, which implicitly captures the current teamwork situation. ODITS jointly learns the marginal utility function and the latent variable by two learning objectives in an end-to-end manner. ODITS sample the teammates' policies $\pi_{j}^{-i}$ from $\{ \pi_{j}^{-i} \}_{j=1,2,\dots, J}^{tr}$ and sample data $D_k = \{ (s_t, a_{t}^{i}, a_{t}^{-i},r_t)  \}_{t=1, \dots, T}$ using the ad hoc agent's policy $\pi^i$ and $\pi^{-i}$ and then add $D_k$ into $\mathcal{D}$. In the training phase, ODITS sample one trajectory $D \sim \mathcal{D}$, compute $(\mu_{c_{t}^{i}}, \sigma_{c_{t}^{i}}) = f(s_t, \mathbf{a}_{t}^{-i})$, compute $(\mu_{z_{t}^{i}}, \sigma_{z_{t}^{i}}) = f^* (b_{t}^i)$, compute $u_{t}^i (r_{t}^i, a_{t}^i; z_{t}^i)$, compute
\begin{equation}
    \mathcal{L}_Q = \mathbb{E}_{(u_{t}^i, c_{t}^i, r_t) \sim D} \big[  [r_t + \gamma \max_{a_{t+1}^i} \Bar{G}(u_{t+1}^i,c_{t+1}^i) - G(u_{t}^i, c_{t}^i)]^2  \big]
\end{equation}
where $\Bar{G}$ is a periodically update target network. And 
\begin{equation}
    \mathcal{L}_{MI} = \mathbb{E}_{(b_{t}^i, s^t, \mathbf{a}_{t}^{-i}) \sim D} [D_{KL}[p(z_{t}^i|b_{t}^i)||q_{\xi}(z_{t}^i|c_{t}^i,b_{t}^i  ) ]]
\end{equation}
where $D$ is the replay buffer, $D_{KL}[\cdot||\cdot]$ is the $KL$ divergence operator. The final objective is to:
\begin{equation}
    \mathcal{L}(\theta) = \mathcal{L}_{Q}(\theta) + \lambda \mathcal{L}_{MI}(\theta_{f^{*}} , \xi)
\end{equation}
The ad hoc agent collects transition data from different training teammates and stores them in the replay buffer $D$. Then, the framework updates all parameters using gradients from the overall loss on samples from $D$. During execution, the ad hoc agent chooses actions that maximize the conditional utility function based on the inferred teamwork situations.\\
\textbf{CSP} learns to cooperate in three steps: (1) It creates a variety of team policies to find different possible coordination schemes. (2) It trains a scheme probing module to represent different teams efficiently by self-supervised team-dynamics reconstruction. (3) It identifies the hidden coordination schemes by clustering the representations and trains a multi-modal meta-policy to adjust to them.

Firstly, CSP train each team policy $\pi_i \in \prod_{train}$ for quality and diversity, train $<E_c, D_c>$ to represent teammates, and $\pi_{sp}$ to prob teammates, then represent teams in $\prod_{train}$ and discover coordination schemes as $k$ clusters. Finally, train $\pi_{meta}$ for generalization by each sub-policy acquiring a unique scheme.\\
\textbf{RPM} generates diverse multi-agent behaviors for self-play to enhance the generalization of MARL. The training episode return $R$ of all agents (with policies $\{ \pi_{\theta}^{i} \}_{i=1}^{N}$) is obtained when they act in the substrate. Then $\{ \pi_{\theta}^{i} \}_{i=1}^{N}$ are saved into RPM by appending agents’ policies into the corresponding memory slot. Then the discretized entry $k$ covers a range of $[k, k + \psi]$, where $\psi > 0$ and is an integer number.
\begin{equation}
    k = \begin{cases} \lfloor R/ \psi \rfloor \times \mathbf{1} \{ (R \quad mod \quad \psi) \neq 0 \} \times \psi, & if \quad R \ge 0 \\   \lfloor R/ \psi \rfloor \times \psi, & otherwise \end{cases}
\end{equation}
where $\mathbf{1} \{ \cdot \}$ is the indicator function, and $\lfloor \cdot \rfloor$ is the floor function.

The RPM memory holds various policies with different performance levels. RPM can choose different policies of different ranks and assign them to each agent in the substrate to produce multi-agent trajectories for training. These diverse multi-agent trajectories can imitate the trajectories created by the interaction with agents with unknown policies in the evaluation scenario. At the start of an episode, we first randomly select N keys with replacement and then randomly select one policy for each key from the corresponding list. All agents’ policies will be changed with the newly selected policies for multi-agent interactions in the substrate, thus creating diverse multi-agent trajectories.
\begin{figure*}[!ht]
\centering
\includegraphics[width=0.8\textwidth]{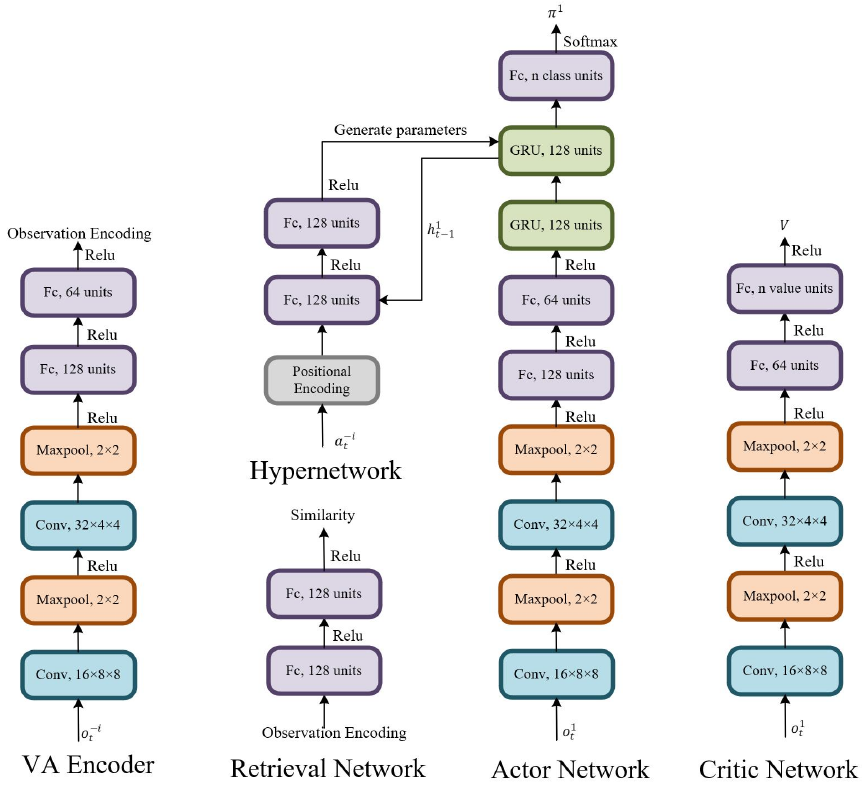} 
\caption{\textbf{Architecture details of MRDG}}
\label{architecture}
\end{figure*}

\section{Cooperative Game}
First, we construct the joint policy from each agent’s point of view in the team, $\pi_{\theta} = (\pi_{\theta^{i}}^{i}(a^{i}|s),\pi_{\theta^{-i}}^{-i}(a^{-i}|s))$. The agent acts at the same time at every stage of the game. The agent aims to get the highest total reward.
\begin{equation}
   max \quad \eta^{i}(\pi_{\theta})=E \big[\sum_{t=1}^{\infty}\gamma^t r^i (s_t, a_t^i, a_t^{-i})\big]
\end{equation}
Let's define the Q function:
\begin{equation}
   Q_{\pi_{\theta}}^{i}=E\big[\sum_{l=0}^{\infty}\gamma^l r^i (s_{t+l}, a_{t+l}, a_{t+l}^{-i})\big]
\end{equation}
The learner agent can only choose its own action, but the reward it gets depends on the actions of others. $\pi_{\theta}(a^{i}, a^{-i}|s) = \pi_{\theta^{i}}^{i}(a^{i}|s)\pi_{\theta^{-i}}^{-i}(a^{-i}|s)$. 

From the learner's point of view, the teammates' strategies $\pi_{\theta^{-i}}^{-i}(a^{-i}|s)$ are unknown. We use the $\rho_{\phi^{-i}}^{-i}(a^{-i}|s, a^{i})$ function to estimate the teammates’ strategies, considering the uncertainty of their actions and the influence of the learner’s actions on them, $\pi_{\theta}(a^{i}, a^{-i}|s) = \pi_{\theta^{i}}^{i}(a^{i}|s)\pi_{\theta^{-i}}^{-i}(a^{-i}|s, a^{-i})$. Our objective for the AHT problem is to achieve the maximum collective reward, instead of the maximum personal reward for each agent, the objective becomes:
\begin{equation}
   argmax_{\theta_{i}, \phi_{-i}} \quad \eta^{i}(\pi_{\theta^{i}}^{i}(a^{i}|s)\rho_{\phi^{-i}}^{-i}(a^{-i}|s, a^{i}))
\end{equation}
We assume that both the learner and teammates are updated towards the overall goal, and we get:
\begin{equation}
\begin{aligned}
   \eta = & \int_s \int_{a^{i}} \\ & \pi_{\theta^{i}}^{i}(a^{i}|s)\int_{a^{t}\in a^{-i}} \pi_{\theta^{-i}}^{-i}(a^{-i}|s, a^{i}) Q^{i}(s, a^{i}, a^{t}) da^{t}da^{i}ds
\end{aligned}
\end{equation}
The gradient update for each agent can be expressed as follows:
\begin{equation}
\begin{aligned}
   \nabla_{\theta^i} \eta^i = & E_{s \sim p, a^{i} \sim \pi^{i}} [\nabla_{\theta^i}log     \pi_{\theta^{i}}^{i}(a^{i}|s) \\ & \int_{a^{-i}} \pi_{\theta^{-i}}^{-i}(a^{-i}|s, a^{i})Q^{i}(s, a^{i}, a^{-i}) da^{-i}]
\end{aligned}
\end{equation}
To obtain the optimal solution of team revenue, it is important to determine the teammate’s strategy. We use the approximate value $\rho_{\phi^{-i}}^{-i}(a^{-i}|s, a^{i}))$ to replace the teammate’s strategy and aim to make the approximate value infinitely close to the teammate’s strategy. This will maximize the team’s benefits. To achieve this goal, MRDG explicitly models the teammates so that the learner can adapt to the current teammates.

\section{Architectures}
We first introduce the network structure and hyperparameters of each module in MRDG as shown in Figure~\ref{architecture}. \\
\subsection{VA Encoder.} The VA Encoder is used to encode the observation information of the learner and teammates. The network structure of VA Encoder is similar to the first few layers of the learner’s strategy network. Two layers of CNN are used to encode the observation image. The first layer of CNN has 16 output channels, and the size of the convolution kernel is 8×8, the second layer of CNN has 32 output channels, and the size of the convolution kernel is 4×4, both of which use the ReLu activation function. The output of the two layers of CNN enters the next two FC layers with 128 and 64 respectively. The FC layer of a unit obtains the observation coding information of the learner and teammates observing viewpoint alignment.\\
\subsection{Retrieval Network.} The main function of the Retrieval Network is to compare the similarity between the current observation and the information in the Episodic memory. We use two FC layers with 128 neural units to model the Retrieval Network module. The output of both layers of FC uses the ReLu activation function. This module receives the current observation and Episodic memory as input and outputs the most similar top $k$ observations in Episodic memory observed action value.\\
\subsection{Hypernetwork.} The main function of Hypernetwork is to generate the network parameters of the second layer of GRU in the learner strategy network. We use a two-layer FC network to represent it. Each layer of the network has 128 neural units, and each layer of FC uses the ReLu activation function. Hypernet receives the teammate action value integrated by Positional Encoding and the output of the last hidden layer information of the second layer GRU in the learner strategy network as input and outputs the network parameters of the GRU.\\
\subsection{The Learner Policy Network.} For the decision-making network parameters of the learner, we use the actor-critic structure. For the observation information coding part, the network structure of actor and critic is the same, with two layers of CNN. The first layer of CNN has 16 output channels, and the convolution kernel size is 8×8. After passing through the 2×2 Maxpool layer, the ReLu activation function is used. The second layer of CNN has 32 output channels, and the size of the convolution kernel is 4×4. After passing through the 2×2 Maxpool layer, the same ReLu activation function is used. The information after the Actor Network is input to two layers of FC layers with 128 and 64 neural units respectively. After using the ReLu activation function, it is input into two layers of GRU neural network to retain the historical information of Markov decision. Then it passes through a layer with the learner’s FC layer with the size of the action space and uses the Softmax activation function to obtain the learner’s policy network. The information after the Critic Network is input to two FC layers with 64 and 1 neurons respectively. After using the ReLu activation function, we obtain the $V$ value of the current strategy, which is used to judge whether the current strategy is good or bad.

\printcredits

\bibliographystyle{model1-num}

\bibliography{ref_file}





\end{document}